\documentclass[12pt]{elsarticle}
\usepackage{ifpdf,amssymb,amsmath,epsfig,bm,pifont}
\usepackage{graphics}
\usepackage{graphicx}
\usepackage{tabularx}
\usepackage{multirow}
\usepackage{dcolumn}
  \graphicspath{{./figs/}}
   \pdfoutput=1
\usepackage{color}

\def\be{\begin{equation}}
\def\ee{\end{equation}}
\def\bea{\begin{eqnarray}}
\def\eea{\end{eqnarray}}
\def\bear{\begin{array}}
\def\eear{\end{array}}
\def\bes{\begin{subequations}}
\def\ees{\end{subequations}}
\newcommand{\MSbar}{\overline{\rm MS}}  
\newcommand{\m}{{\overline m}}
\newcommand{\A}{{\mathcal{A}}}
\newcommand{\tA}{{\widetilde {\mathcal{A}}}}

\newcommand{\tk}{{\widetilde k}}

\newcommand{\bL}{{\overline{\Lambda}}}

\renewcommand{\thefootnote}{\fnsymbol{footnote}}

\biboptions{compress}

\begin{document}
\begin{frontmatter}

\title{\vskip-3cm{\baselineskip14pt
\centerline{\normalsize\hfill USM-TH-328}
}\vskip.7cm
\texttt{anQCD}: a \texttt{Mathematica}  
package for calculations in general analytic QCD models
\vskip.3cm
}
 \author[utfsm]{C\'esar Ayala}
\ead{c.ayala86@gmail.com}

 \author[utfsm]{Gorazd Cveti\v{c}}
\ead{gorazd.cvetic@usm.cl}

\address[utfsm]{Department of Physics, Universidad T\'ecnica Federico Santa Mar\'ia,\\
                Casilla 110-V, Valpara\'iso, Chile\\}

\date{}

\begin{abstract}
\noindent
We provide a \texttt{Mathematica} package 
that evaluates the QCD analytic couplings (in the complex domain)
$\mathcal{A}_{\nu}(Q^2)$, which are analytic analogs of the powers
$a(Q^2)^{\nu}$ of the underlying perturbative QCD (pQCD) coupling
$a(Q^2) \equiv \alpha_s(Q^2)/\pi$,
in three analytic QCD models (anQCD):
Fractional Analytic Perturbation Theory (FAPT), 
Two-delta analytic QCD (2$\delta$anQCD), 
and Massive Perturbation Theory (MPT).
The analytic (holomorphic) running couplings $\mathcal{A}_{\nu}(Q^2)$, 
in contrast to the corresponding pQCD expressions $a(Q^2)^{\nu}$,  
reflect correctly the analytic properties of the spacelike observables
${\cal D}(Q^2)$ in the complex $Q^2$ plane as dictated by the
general principles of quantum field theory. They are thus 
more suited for evaluations of such physical quantities,
especially at low momenta $|Q^2| \sim 1 \ {\rm GeV}^2$. 

\vspace{.2cm}

\noindent
PACS numbers: 12.38.Bx, 11.15.Bt, 11.10.Hi, 11.55.Fv

\end{abstract}

\end{frontmatter}

\thispagestyle{empty}
 \newpage
  \setcounter{page}{1}

\renewcommand{\thefootnote}{\arabic{footnote}}
 \setcounter{footnote}{0}

\section*{Program Summary}
\begin{itemize}

\item[]\textit{Title of program:}
  \texttt{anQCD}

\item[]\textit{The main program (}\texttt{anQCD.m}\textit{) and supplementary modules (}\texttt{Li\_\_nu.m} \textit{and} \texttt{s0r.m}\textit{), 
and the zipped file containing all three files (}\texttt{anQCD\_Mathematica.zip}\textit{), available from the web page:}\\  \texttt{gcvetic.usm.cl}

\item[]\textit{Computer for which the program is designed and others on which it
    is operable:}
  Any work-station or PC where \texttt{Mathematica} is running.

\item[]\textit{Operating system or monitor under which the program has been
    tested:}
  Operating system Linux and Mac OS X, software Mathematica 9.0.1, 10.0.1 and 10.0.2

\item[]\textit{No. of bytes in distributed program including test data etc.:}\\
  63 kB (main module \texttt{anQCD.m}), 2 kB (supplementary module \texttt{Li\_\_nu.m}), 18 kB (supplementary module \texttt{s0r.m});

\item[]\textit{Distribution format:}
  ASCII
\item[]\textit{Keywords:}
  Analyticity,
  Fractional Analytic Perturbation Theory,
  Two-delta analytic QCD model,
  Massive Perturbation Theory,
  Perturbative QCD,
  Renormalization group evolution.

\item[]\textit{Nature of the physical problem:}
  Evaluation of the values for analytic couplings 
  $\A_{\nu}(Q^2;N_f)$ in analytic QCD [the analytic analog of the
  power $(\alpha_s(Q^2;N_f)/\pi)^{\nu}$] 
  based on the dispersion relation; $\A_{\nu}$ represents a physical 
  (holomorphic) function in the plane of complex squared 
  momenta $-q^2 \equiv Q^2$. In \texttt{anQCD.m} we collect the formulas 
  for three different analytic models depending on the 
  energy scale, $Q^2$, number of flavors $N_f$, the QCD scale
  ${\overline {\Lambda}}_{N_f}$, 
  and the (nonpower) index $\nu$. The considered models are:
  Analytic Perturbation theory (APT), Two-delta 
  analytic QCD (2$\delta$anQCD) and Massive Perturbation Theory (MPT). 

\item[]\textit{Method of solution:}
  \texttt{anQCD} uses \texttt{Mathematica} functions to perform
  numerical integration of spectral function for each analytic 
  model, in order to obtain the corresponding analytic images
  $\A_{\nu}(Q^2)$ via dispersion relation.

\item[]\textit{Restrictions on the complexity of the problem:}
  It could be that for an unphysical choice of the input parameters
  the results are meaningless.

\item[]\textit{Typical running time:}
  For all operations the running time does not exceed a few seconds.

\end{itemize}

\newpage

\section{Introduction}
 \label{sec:intro}
The perturbative approach to QCD (pQCD) works
well for evaluations of physical quantities at high
momentum transfer ($|q^2| \gtrsim 10^1 \ {\rm GeV}^2$).
However, it is unreliable at low momenta ($|q^2| \sim 1 \ {\rm GeV}^2$),
the principal reason for this being the existence of
singularities of the pQCD
coupling parameter $a(Q^2) \equiv \alpha_s(Q^2)/\pi$ 
(where $Q^2 \equiv - q^2$)
at such complex spacelike momenta $Q^2$:
$|Q^2| \lesssim  1 \ {\rm GeV}^2$ and $Q^2 \not< 0$.
These (Landau) singularities
reappear in evaluations of the spacelike observables ${\cal D}(Q^2)$
for small $|Q^2|$. For example, if  ${\cal D}(Q^2)$ is dominated
by the leading-twist term of dimension zero,
its evaluated expression is $f(a(\kappa Q^2))$ where
$f$ is a (truncated) power series in $a(\kappa Q^2)$ and the positive
$\kappa$ ($\sim 1$) is the renormalization scale parameter. 
Hence $f(a(\kappa Q^2))$ has the same region of singularities as $a(\kappa Q^2)$.
This does not reflect correctly the true analyticity structure of the
spacelike observable ${\cal D}(Q^2)$. Such an observable must be,
by the general principles of the local 
quantum field theory \cite{BS,Oehme},
a holomorphic (analytic) function in the complex $Q^2$ plane except
on parts of the negative semiaxis where it has a cut; i.e., 
analyticity for $Q^2 \in \mathbb{C} \backslash (-\infty, 0]$.
Therefore, the coupling parameter $\mathcal{A}_1(Q^2)$, 
that is to be used instead of $a(Q^2)$ to evaluate the 
spacelike observables ${\cal D}(Q^2)$, should have qualitatively
the same analyticity properties, i.e., $\mathcal{A}_1(Q^2)$ should be
a holomorphic function for  $Q^2 \in \mathbb{C} \backslash (-\infty, 0]$.
Such an analytic function $\mathcal{A}_1(Q^2)$
defines what is called analytic QCD (anQCD) model.

The finiteness of the QCD coupling in the infrared regime and,
in general, the holomorphic behavior of it in the $Q^2$ complex plane,
are suggested by various independent lines of research in QCD,
among them: by the Gribov-Zwanziger approach \cite{Gribov}; by
analyses of Dyson-Schwinger equations in QCD \cite{DSE1,DSE2}
and by other functional methods \cite{STQ,FRG}; by
lattice calculations \cite{lattice}; by models using the AdS/CFT
correspondence modified by a dilaton backgound \cite{AdS};
in various other approaches such
as those in Refs.~\cite{Simonov,BadKuz,BKS,KKSh,Deur,Court}.

The first anQCD model, constructed explicitly in the aforementioned sense, 
is the Analytic Perturbation Theory (APT) of
Shirkov, Solovtsov {\it et al.\/}~\cite{ShS,MS96,ShS98,Sh}. 
The underlying pQCD discontinuity function
$\rho_1^{\rm (pt)}(\sigma) \equiv {\rm Im} a(Q^2=-\sigma - i \epsilon)$
was kept unchanged on the entire negative axis in the $Q^2$-plane, 
i.e., ${\rm Im} \mathcal{A}_1^{\rm (APT)}(-\sigma - i \epsilon) =
\rho_1^{\rm (pt)}(\sigma)$ for all $\sigma \geq 0$. On the other hand, the
Landau discontinuity region (at $-\Lambda_{\rm Lan.}^2 \leq \sigma < 0$)
was eliminated, i.e.,
${\rm Im} \mathcal{A}_1^{\rm (APT)}(-\sigma - i \epsilon) = 0$ for $\sigma < 0$.
The resulting coupling $\mathcal{A}_1^{\rm (APT)}(Q^2)$
for $Q^2 \in \mathbb{C} \backslash (-\infty, 0]$ was then obtained by
the use of a dispersion relation involving $\rho_1^{\rm (pt)}(\sigma)$ at
$\sigma \geq 0$.
The analogs $\mathcal{A}_n^{\rm (APT)}(Q^2)$ of integer powers $a(Q^2)^n$
were also constructed in the aforementioned works.
An extension to the analogs $\mathcal{A}_{\nu}^{\rm (APT)}(Q^2)$
of noninteger powers $a(Q^2)^{\nu}$ in this model were obtained
and used in the works ~\cite{BMS05,BKS05,BMS06,BMS10}; hence this anQCD model
is also called Fractional APT (FAPT).

Later on, other analytic QCD models were constructed, which fulfill
certain additional physically motivated restrictions, such as 
Refs.~\cite{Nest1,Nest2,Webber,CV1,CV2,Alekseev,1danQCD,2danQCD,MPT}.
Analytic QCD models, as well as related dispersive approaches,
have been used in evaluations of various low-momentum QCD quantities,
cf.~Refs.~\cite{MSS2,MagrGl,mes2,DeRafael,MagrTau,Nest3,anOPE}.
Reviews of the analytic QCD approaches are given in Refs.~\cite{Prosperi,Shirkov,CVrev,Bakulev,BaSh,Ste}.

In addition to FAPT, we will consider here the Two-delta analytic QCD 
(2$\delta$anQCD)~\cite{2danQCD} 
and Massive Perturbation Theory (MPT)~\cite{MPT}.
The 2$\delta$anQCD model \cite{2danQCD} is similar to
FAPT model in the sense that it
is (partially) based on the underlying pQCD coupling $a(Q^2)$:
${\rm Im} \mathcal{A}_1^{\rm (2 \delta)}(-\sigma - i \epsilon) = 
\rho_1^{\rm (pt)}(\sigma)$ for large enough $\sigma \geq M_0^2$ 
(where $M_0 \sim 1$ GeV is a ``pQCD-onset'' scale). 
On the other hand, in
the (otherwise unknown) low-$\sigma$ regime, $0 < \sigma < M_0^2$,
the behavior of the discontinuity function 
$\rho_1(\sigma) \equiv 
{\rm Im} \mathcal{A}_1^{\rm (2 \delta)}(Q^2=-\sigma - i \epsilon)$
is parametrized by two positive delta functions. The coupling
$\mathcal{A}_1^{\rm (2 \delta)}(Q^2)$ is then obtained by the use of
a dispersion relation involving $\rho_1(\sigma)$. The parameters
for the delta functions and $M_0$ are determined by requiring that
the model effectively merges with the pQCD for large 
$|Q^2| >  \Lambda^2$ (where $\Lambda^2 \sim 0.1 \ {\rm GeV}^2$),
and by requiring that the model reproduce the experimentally
determined value $r_{\tau}=0.203$ of the $\tau$ lepton
semihadronic nonstrange $V+A$ decay rate ratio.
On the other hand,
Massive Perturbation Theory (MPT)~\cite{MPT} is defined via
the identity $\mathcal{A}_{1}(Q^2) = a(Q^2+m^2_{\rm gl})$, where 
$m_{\rm gl} \sim 1$ GeV
is an effective dynamical gluon mass.

In general anQCD models, such as 2$\delta$anQCD or MPT, the formalism
for construction of analytic analogs $ \mathcal{A}_{\nu}(Q^2)$ of
the powers $a(Q^2)^{\nu}$ was formulated in Refs.~\cite{CV1,CV2} for
the case of integer index $\nu$, and in Ref.~\cite{GCAK}
for general (noninteger) index $\nu$. Generally we have
$\mathcal{A}_{\nu} \not= (\mathcal{A}_1)^{\nu}$.

Presently, there exist programs for numerical evaluation of
the APT and ``massive'' APT (MAPT) \cite{BK1},
and of FAPT couplings \cite{BK2}. The purpose of this work is
to offer an extended program in \texttt{Mathematica} which numerically
evaluates the couplings in FAPT, 2$\delta$anQCD and in MPT,
in order to correctly evaluate (truncated) perturbation series of
physical quantities in these anQCD models. Our program evaluates the
FAPT couplings in a similar way as the program of Ref.~\cite{BK2};
but the part of our program which evaluates the  
2$\delta$anQCD and MPT couplings is new.

We summarize in Sec.~\ref{sec:pQCD} the calculation of the
running coupling of the underlying pQCD, the threshold matching,
and the corresponding QCD scales ${\overline \Lambda}_{N_f}$. 
In Sec.~\ref{sec:anQCD} we present a general method for calculation of the
analytic analogs $\A_{\nu}(Q^2)$ of powers $a(Q^2)^{\nu}$ in anQCD models,
and a description of the three mentioned anQCD models: 
FAPT, 2$\delta$anQCD (with new extension for $N_f \geq 4$), and MPT.
In addition, curves of some of the resulting couplings as a function
of $Q^2$, at positive $Q^2$, are presented.
Finally, in Sec.~\ref{sec:pract} we present some practical aspects
and the main procedures of the calculational
program, as well as some specific examples. More
detailed definitions of the procedures are included in \ref{sec:def}. 

\section{Running coupling in the underlying perturbative QCD}
 \label{sec:pQCD}
\subsection{Running coupling at fixed $N_f$}
\label{sec:fixedNf}

The differential equation that defines the beta function and therefore the running coupling in perturbative QCD (pQCD) is given by the renormalization group equation (RGE, at renormalization scale $\mu^2=Q^2$)
\be
\beta(a(Q^2))=Q^2\frac{\partial a(Q^2)}{\partial Q^2}=-\sum_{j=2}^\infty \beta_{j-2}(N_f)a^j(Q^2),
\label{beta}
\ee
with the notation: $a(Q^2)\equiv\alpha_s(Q^2)/\pi=g_s(Q^2)^2/(4\pi^2)$ and $N_f$ is the number of active quarks flavors. 
The first two beta coefficients ($\beta_0$ and $\beta_1$, \cite{bet0,bet1}) are scheme independent, i.e., they are universal in the mass independent renormalization schemes
\bea
\label{betacoeff01}
\beta_0(N_f)&=&\frac{1}{4}\left(11-\frac{2}{3}N_f \right), \qquad
\beta_1(N_f)=\frac{1}{16}\left(102-\frac{38}{3}N_f \right).
\eea
The next coefficients ($\beta_2, \beta_3,\ldots$) are scheme dependent;
in fact, they define the renormalization scheme \cite{Stevenson}. In
the $\MSbar$ scheme, $\beta_2$ and $\beta_3$ are known \cite{bet2,bet3}
\bes
\label{betacoeff23}
\bea
{\overline \beta}_2(N_f)&=&\frac{1}{64}\left(\frac{2857}{2}-\frac{5033}{18}N_f+\frac{325}{54}N_f^2 \right),\\
{\overline \beta}_3(N_f)&=&\frac{1}{256}\left[\left(\frac{149753}{6}+3564\zeta_3\right)-\left(\frac{1078361}{162}+\frac{6508}{27}\zeta_3\right)N_f
\right.
\nonumber\\
&&
\left.
+\left(\frac{50065}{162}+\frac{6472}{81}\zeta_3\right)N_f^2+\frac{1093}{729}N_f^3 \right]
\eea
\ees
where $\zeta_\nu$ is the Riemann zeta function, in particular $\zeta_3\simeq1.202057$.

The beta function on the right-hand side of Eq.~(\ref{beta}) is usually 
approximated as a truncated perturbation series of coupling $a$. The resulting differential equation for $a$ is solved, either analytically (if possible) or numerically.
For example, the one-loop order equation can be integrated explicitly,
giving the well known solution
\be
a(Q^2)=\frac{1}{\beta_0 {\rm ln}(Q^2/{\overline \Lambda}^2)}, \qquad {\overline \Lambda}^2=\mu^2 e^{-1/(\beta_0 a(\mu^2))}.
\label{1LpQCD}
\ee
One way to solve the RGE at the two-loop level is to iterate with respect to the one-loop formula. This gives us an approximate coupling as an expansion in powers of $L^{-1}$, where $L\equiv {\rm ln}(Q^2/{\overline \Lambda}^2)$. If we truncate at $L^{-2}$, we obtain
\be
a^{(2,L2)}(Q^2)=\frac{1}{\beta_0 L}\left(1-\frac{\beta_1}{\beta_0^2}\frac{{\rm ln}(L)}{L} \right) \ .
\label{2LpQCDapp}
\ee
The iterative method can be performed at any loop level.
For example, when truncating the expansion of the $M$-loop coupling
at $L^{-{\cal N}} \equiv 1/\ln^{\cal N}(Q^2/{\overline \Lambda}^2)$, we obtain\footnote{
The superscript notation $(M,L{\cal N})$ in Eq.~(\ref{NLpQCDapp}) means that the expansion is truncated at $1/L^{\cal N}$, and that $M$-loop $\beta$-function is taken, i.e., $\beta_j=0$ for $j \geq M$. For consistency reasons, we must have ${\cal N} \geq M$. In practice, the expansion gives us expression which, for $Q^2 > {\overline \Lambda}^2$, tends toward the exact $M$-loop coupling $a^{(M)}(Q^2)$ when ${\cal N} \to \infty$
(i.e., ${\cal N} \gg M$).} 
\bea
a^{(M,L{\cal N})}(Q^2)&=&\frac{1}{\beta_0 L}\left\{
1-\frac{\beta_1}{\beta_0^2}\frac{{\rm ln}(L)}{L}
+\frac{1}{\beta_0^2 L^2}\left[\frac{\beta_1^2}{\beta_0^2}({\rm ln}^2(L)-{\rm ln}(L)-1)+\frac{\beta_2}{\beta_0} \right] +
\right.
\nonumber\\
&&
\left.
\frac{1}{\beta_0^3 L^3}\left[\frac{\beta_1^3}{\beta_0^3}\left(-{\rm ln}^3(L)+\frac{5}{2}{\rm ln}^2(L)+2{\rm ln}(L)-\frac{1}{2} \right)
-3\frac{\beta_1\beta_2}{\beta_0^2}{\rm ln}(L)+\frac{\beta_3}{2\beta_0} \right]
+ \right.
\nonumber\\
&&
\left.
\frac{1}{\beta_0^{{\cal N}-1} L^{{\cal N}-1}} \left[
\frac{\beta_1^{{\cal N}-1}}{\beta_0^{{\cal N}-1}} (-1)^{{\cal N}-1} \ln^{{\cal N}-1} L + \ldots \right] \right\} \ .
\label{NLpQCDapp}
\eea
There is a way to find the two-loop coupling as a solution of RGE exactly. The two-loop RGE leads to  a transcendental equation. Namely, integrating ~(\ref{beta}), with $\beta_k=0$ for $(k=2,3,\ldots)$,  we have
\be
\int_{a(\mu^2)}^{a(Q^2)}\frac{da}{a^2\left(1+\frac{\beta_1}{\beta_0}a \right)}=-\beta_0\int_0^{\frac{1}{2}{\rm ln}(Q^2/\mu^2)}d{\rm ln}(Q'^2/\mu^2).
\ee
So, the transcendental equation gets the form
\be
{\rm ln}(Q^2/\mu^2)=C+\frac{1}{\beta_0a(Q^2)}+\frac{\beta_1}{\beta_0^2}{\rm ln}(a(Q^2))-\frac{\beta_1}{\beta_0^2}{\rm ln}\left(1+\frac{\beta_1}{\beta_0}a(Q^2) \right),
\ee
where $C$ contains the coupling $a(\mu^2)$.\\
A new invariant mass parameter $\Lambda$ can be introduced, given by
 \bea
 {\rm ln}(Q^2/\Lambda^2)=\frac{1}{\beta_0a(Q^2)}-\frac{\beta_1}{\beta_0^2}{\rm ln}\left(\frac{\beta_1}{\beta_0^2}+\frac{1}{\beta_0a(Q^2)} \right),
 \nonumber\\
 \qquad\qquad
 \Lambda^2=\mu^2{\rm exp}\left[C-\frac{\beta_1}{\beta_0^2}{\rm ln}(\beta_0) \right].
 \eea
This relation must be inverted; however, nontrivial
problems related to the singularity structure appear. 
The solution is achieved with the help of the so-called Lambert $W$ function 
defined by
\be
W(z){\rm exp}[W(z)]=z \ .
\label{LambertDef}
\ee
The singularity structure of the Lambert function consists of an 
infinite number of branches; it satisfies the following symmetry relation: 
$W_{-n}^\ast(y^\ast)=W_n(y)$.\\
With this function the solution to the coupling is \cite{MagrGl,2lLambert}
\be
a^{(2)}(Q^2)=-\frac{1}{c_1}\frac{1}{1+W_{\mp1}(z_{\pm})},
\label{a2lLambert}
\ee
where $c_1=\beta_1/\beta_0$, $Q^2=|Q^2|{\rm e}^{i\phi}$, and the upper sign refers to the case $0\leq\phi\leq+\pi$, the lower sign to $-\pi\leq\phi\leq0$, and
\be
z_{\pm}=\frac{1}{c_1{\rm e}}\left(\frac{|Q^2|}{\Lambda^2} \right)^{-\beta_0/c_1}{\rm exp}\left[i\left(\pm\pi-\frac{\beta_0}{c_1}\phi \right) \right].
\label{zina2l}
\ee

\begin{figure}[htb] 
\begin{minipage}[b]{.5\linewidth}
\centering\includegraphics[width=75mm]{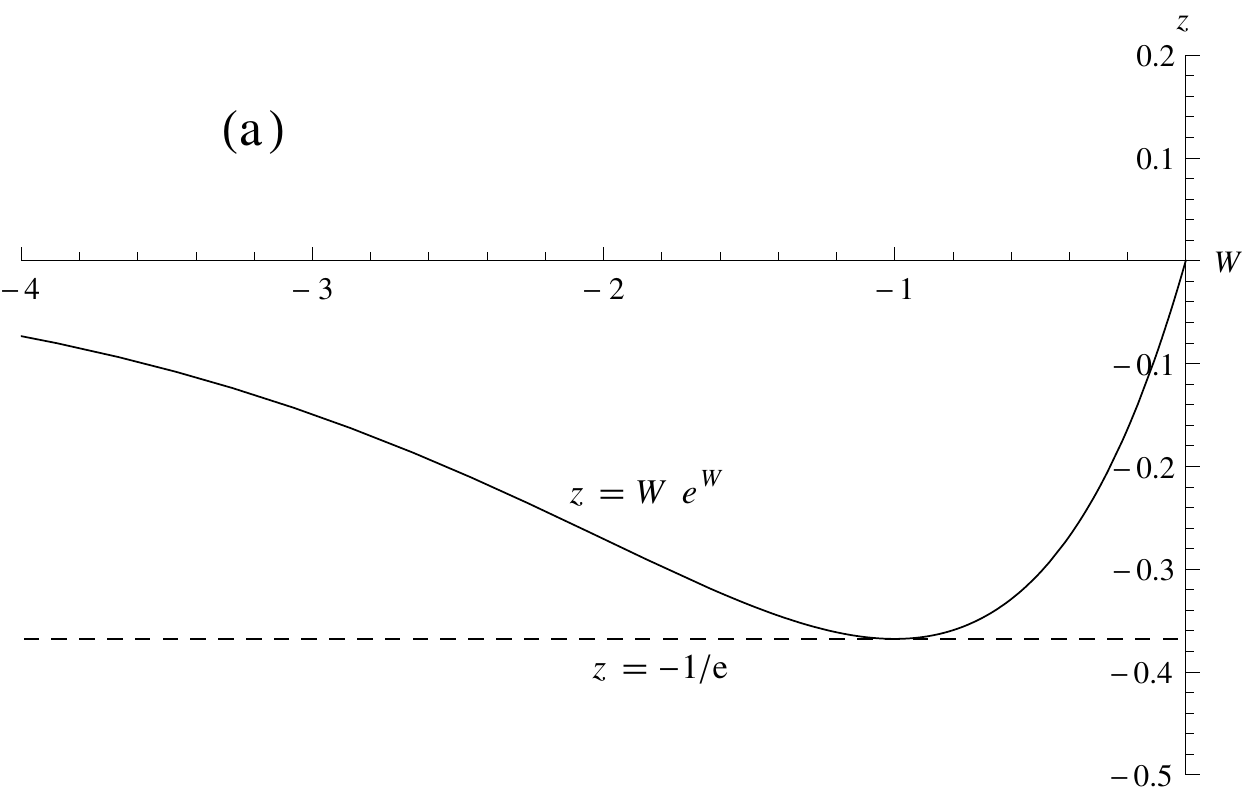}
\end{minipage}
\begin{minipage}[b]{.5\linewidth}
\centering\includegraphics[width=75mm]{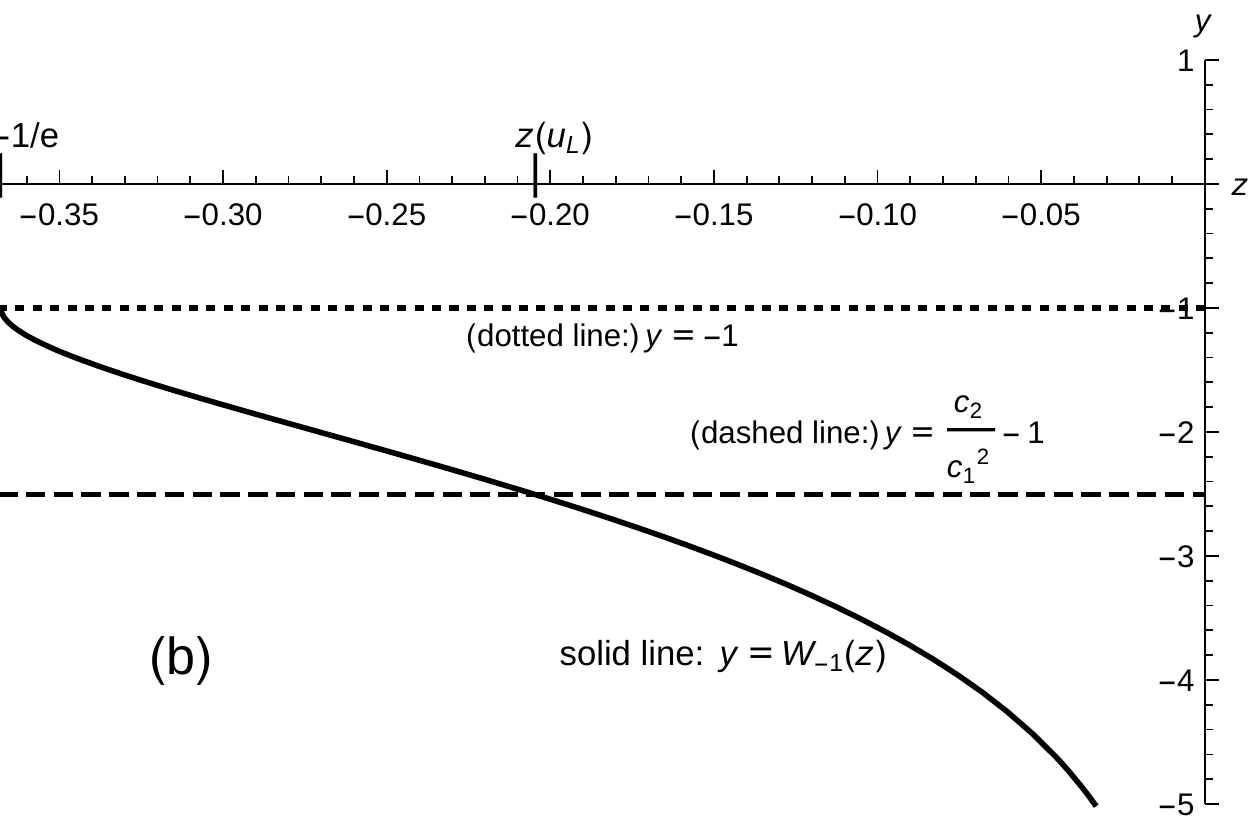}
\end{minipage}
\vspace{-0.4cm}
 \caption{\footnotesize (a) The defining relation $z = W e^W$ for
the Lambert function $W(z)$, for $-1/e < z < 0$; 
(b) The branch $W_{-1}(z)$ for the same $z$-interval; when $c_2<0$,
the denominator of Eq.~(\ref{aptexact}) becomes zero at a
$z(u_{\rm L})$ in this interval.}
\label{figACC1}
 \end{figure}

This idea can be extended to the higher loop case, using for the beta function
$\beta(a)$ the form of Pad{\'e} $[3/1](a)$
\bes
\label{betaPad}
\bea
\lefteqn{
\beta_{[3/1]}(a) =
-\beta_0 a^2(Q^2) \frac{1+(c_1-c_2/c_1)a(Q^2)}{1-(c_2/c_1)a(Q^2)}
}
\label{betaina2l}
\\
& = & -\beta_0 a^2(Q^2) \left[ 1 + c_1 a(Q^2) + c_2 a(Q^2)^2 + \frac{c_2^2}{c_1} a(Q^2)^2 + \frac{c_2^3}{c_1^2} a(Q^2)^3 + \ldots \right],
\label{betaPadexp}
\eea
\ees
where the renormalization scheme parameters are: $\beta_2=\beta_0 c_2$ and
$\beta_j=\beta_0 c_2^{j-1}/c_1^{j-2}$ ($j \geq 3$). We call this
scheme $c_2$-Lambert scheme. When $c_2$ in the beta function
(\ref{betaPad}) is chosen to be in $\MSbar$
scheme, i.e., $c_2={\overline c}_2$ ($={\overline \beta}_2/\beta_0$),
we will refer to this scheme, somewhat loosely, as 3-loop $\MSbar$
in pQCD, FAPT and MPT (the 4-loop coefficient 
$\beta_3= \beta_0 {\overline c}_2^2/c_1$ is not $\MSbar$).
With this, the solution of the coupling to three loops in terms of the Lambert function takes the form \cite{3lLambert}\footnote{In the expression (\ref{zina2l}), the ``Lambert'' scale $\Lambda$ is different from the scale $\overline{\Lambda}$ appearing in the expansion (\ref{NLpQCDapp}). Therefore, as we use the latter as an input, the program relates these two scales by equating Eq.~(\ref{NLpQCDapp}) [with: $\beta_j/\beta_0 \equiv c_j = c_2^{j-1}/c_1^{j-2}$ ($j=2,3,\ldots$), cf.~the expansion (\ref{betaPadexp})] with Eq.~(\ref{aptexact}) at high $Q^2$ ($\sim 10^{10}{\rm GeV}^2$).}
\be
a(Q^2)=-\frac{1}{c_1}\frac{1}{1-c_2/c_1^2+W_{\mp1}(z_{\pm})} \ .
\label{aptexact}
\ee
The Lambert function $W=W(z)$ is defined via the inverse relation~(\ref{LambertDef}), cf.~Fig.~\ref{figACC1}(a).


The two branches $W_{\mp 1}(z)$ of the Lambert function are related
via complex-conjugation $W_{+1}(z^{*}) = W_{-1}(z)^{*}$, 
and the point $z=-1/e$ is the branching point of these functions.
In the interval $-1/e < z < 0$, $W_{-1}(z)$ is a decreasing function of $z$,
cf.~Fig.~\ref{figACC1}(b). When $z \to -0$, the scale $Q^2$ tends
to $Q^2 \to + \infty$, and $W_{-1}(z) \to - \infty$,
this reflecting the asymptotic freedom of $a(Q^2)$ of
Eq.~(\ref{aptexact}). 

The coupling (\ref{aptexact}) with the $\MSbar$ value
$c_2={\overline c}_2(N_f) 
\equiv {\overline \beta}_2(N_f)/{\overline \beta}_0(N_f)$ 
will be the underlying pQCD coupling in those analytic models which we call:
3-loop ${\rm FAPT}_{N_f}$, 3-loop global FAPT,\footnote{
For the details of the definition of the ($\MSbar$)
3-loop global FAPT, see Secs.~\ref{sec:thr} and \ref{sec:FAPT}.}
and 3-loop ${\rm MPT}_{N_f}$.
In 2$\delta$anQCD, the underlying pQCD coupling will also be that of 
Eq.~(\ref{aptexact}), but with the scheme parameter $c_2$ in the interval
$-5.6 < c_2 < -2$, cf.~Table \ref{2dtab} later (with $c_2=-4.9$ 
being the preferred illustrative value).

\subsection{Thresholds and global coupling}
  \label{sec:thr}

We note that the dependence on the number of effective quark flavors ($N_f$) 
is in the beta coefficients~(\ref{betacoeff01})-(\ref{betacoeff23}). 
We use the following notations:
the $N_f$'th quark flavor has the $\MSbar$  mass 
$m_{N_f} \equiv \m_q \equiv \m_q(\m_q)$, where $q=c,b,t$ for $N_f=4,5,6$, 
respectively (we consider $\m_u, \m_d, \m_s \approx 0$). ${\rm QCD}_{N_f}$ is
applied, in principle, at the scales $\mu \equiv \sqrt{|Q^2|}$ such that
$m_{N_f} \ll \mu \ll m_{N_f+1}$; in practice, it is applied at
$\mu$'s such that $m_{N_f} \lesssim \mu \lesssim m_{N_f+1}$.
If the threshold scale is chosen to be $Q^2_{\rm thr} = m_{N_f}^2$,
the one-loop quark threshold condition is the continuity of the
coupling $a(Q^2)$ there; i.e., 
at $Q^2 = m_{N_f}^2$ we have for $a(Q^2, {\overline \Lambda}^2, N_f)$
\be
a(m_{N_f}^2, {\overline \Lambda}^2_{N_f-1}, N_f-1) = 
a(m_{N_f}^2, {\overline \Lambda}^2_{N_f}, N_f) \ .
\label{contthr}
\ee
At a higher loop level, a noncontinuous matching has to be performed between the couplings in the effective theories ${\rm QCD}_{N_f}$ and ${\rm QCD}_{N_f-1}$.
If the coupling runs according to the $N$-loop $\MSbar$ beta function, 
the $(N-1)$-loop matching condition should be used.  
According to the results of Ref.~\cite{thr4l}, the 3-loop matching 
condition (for the case of 4-loop $\MSbar$ RGE running) has the form
\begin{eqnarray}
\label{msb}
a^\prime &=&a-a^2 \frac{\ell_h}{6}
+a^3 \left(\frac{\ell_h^2}{36}-\frac{19}{24}\ell_h+ {\widetilde c}_2\right)
+a^4\left[-\frac{\ell_h^3}{216}
\right.\nonumber\\
&-&\left.\vphantom{\frac{\ell_h^3}{216}}
\frac{131}{576}\ell_h^2+\frac{\ell_h}{1728} \left( -6793+281 (N_f-1) \right)+ {\widetilde c}_3\right],
\end{eqnarray}
where: $\ell_h=\ln[\mu_{N_f}^2/{\m}_q^2]$; 
$a^\prime=a(\mu_{N_f}^2;N_f-1)$ and $a=a(\mu_{N_f}^2; N_f)$ in $\MSbar$; and
\begin{equation}
\label{cms}
{\widetilde c}_2=\frac{11}{72},\quad
{\widetilde c}_3=-\frac{82043}{27648}\zeta_3+\frac{564731}{124416}
-\frac{2633}{31104} (N_f-1) \ .
\end{equation}
The threshold scale is $\mu^{(N_f)} = \kappa \m_q$
($\ell_h = 2 \ln \kappa$), where $q=c,b,t$ for $N_f=4,5,6$, respectively; and usually $1 \leq \kappa \leq 3$ is taken.\footnote{
For the evaluation of $a(Q^2)$ at a complex $Q^2$, the $N_f$ value assigned
is determined by $(\kappa m_{N_f})^2 < |Q^2| < (\kappa m_{N_f+1})^2$, i.e.,
with such $N_f$ we have $a(Q^2) = a(Q^2;N_f)$.}

\begin{table}
\caption{Comparison between different values of the scales 
${\overline \Lambda}_{N_f}$ (in MeV)
for various $N_f$, and the values of the $\MSbar$ 
coupling ${\overline a}$ at various thresholds: 
(a) the first line is for the 3-loop threshold matching 
(\ref{msb}) at thresholds $2 \m_q$ and 4-loop RGE-running in $\MSbar$ scheme
(${\overline \beta}_j=0$ for $j \geq 4$); 
(b) the second line is for 1-loop threshold matching at $2 \m_q$ and 2-loop RGE running;
(c) as the case (b), but with $\kappa=1$, i.e.,  the continuous conditions (\ref{contthr}) at thresholds $\m_q$.
In all cases, the expansions (\ref{NLpQCDapp}) with ${\cal N}=8$, and 
the world average value $\alpha_s(M_Z^2,\MSbar)=0.1184$ \cite{PDG2012} are used.}
\begin{center}
\centering
\resizebox{\columnwidth}{!}{%
\begin{tabular}{|l|c|c|c|c|c|c|c|}
\hline
\multirow{2}{5mm}{Method} & \multicolumn{4}{ |c|}{${\overline \Lambda}_{N_f}$} & \multicolumn{3}{ |c| }{${\overline a}(N_f)$ (${\overline a}(N_f-1)$)}
\\
\cline{2-8}
 & ${\overline \Lambda}_{6}$ &  ${\overline \Lambda}_{5}$ &  ${\overline \Lambda}_{4}$ & ${\overline \Lambda}_{3}$ & $N_f=6$ & $N_f=5$ & $N_f=4$ 
\\
\hline
4/3-loop, $\kappa=2$ & 90.6 & 213.3 & 297.0 & 341.8 & 0.03187(0.03161) & 0.05948(0.05842) & 0.08706(0.08446)
\\
\hline
2/1-loop, $\kappa=2$ & 89.7 & 216.7 & 312.6 & 375.3 & 0.03185(0.03162) & 0.05934(0.05852) & 0.08650(0.08477)
\\
\hline
2/1-loop, $\kappa=1$ & 90.7 & 216.7 & 308.1 & 361.8 & 0.03465(0.03465) & 0.07154(0.07154) & 0.12061(0.12061)
\\
\hline
\end{tabular}
}
\end{center}
\label{bLtab}
\end{table}
In Table \ref{bLtab}, we present the results for various scales 
${\overline \Lambda}_{N_f}$ in pQCD, for the case of the 4-loop RGE 
running in $\MSbar$ scheme and the corresponding 3-loop threshold 
matching with $\kappa=2$ [thresholds at $Q=\kappa \m_q$], i.e., 
the 4/3-loop case; and for the 2-loop RGE running and 1-loop 
threshold matching with $\kappa=2$ and $\kappa=1$, 
i.e., the 2/1-loop case. For the 
starting value in the numerical integration of the RGE, we used the 
present world average value $a(M_Z^2;N_f=5)=0.1184/\pi$ \cite{PDG2012} 
in $\MSbar$. In all cases, the values of ${\overline \Lambda}_{N_f}$ 
were determined by equating the numerically obtained (``exact'') values 
of $a(Q^2)$ with those of the expansion (\ref{NLpQCDapp}) with ${\cal N}=8$;
the matchings for $N_f=5,4,3$ were made at the corresponding positive 
maximal values of the $N_f$-range, i.e., at $Q^2 = (\kappa \m_q)^2$, 
where $\m_q=\m_t, \m_b, \m_c$ for $N_f=5,4,3$, respectively. 
The used values of the $\MSbar$ masses $\m_q \equiv \m_q(\m_q)$ 
were: $1.27$ GeV \cite{PDG2012}, $4.2$ GeV \cite{ACP}, $163$ GeV (cf., e.g., \cite{CCG}), respectively. 
The value of the scale ${\overline \Lambda}_6$ was determined by equating
the  expansion (\ref{NLpQCDapp}) with the numerical values $a(Q^2;N_f=6)$ at large momenta ($Q \gtrsim 10^3$ GeV). The 4/3-loop results change insignificantly when the threshold matching parameter changes from $\kappa=2$ to $\kappa=1$: ${\overline \Lambda}_3$ value decreases by 1.2 MeV, and ${\overline \Lambda}_4$ value by 0.4 MeV. The 2/1-loop values, however, change significantly when we change $\kappa=2$ to $\kappa=1$: ${\overline \Lambda}_3$ decreases from 375.3 to 361.8 MeV; ${\overline \Lambda}_4$ decreases from 312.6 to 308.1 MeV. In all cases (4/3 and 2/1-loop), the value  ${\overline \Lambda}_5$ is independent of $\kappa$ (because the initial value is at $Q^2=M_Z^2$, i.e., where $N_f=5$); the value of ${\overline \Lambda}_6$ varies insignificantly in the 4/3-loop case, and in the 2/1-loop case it increases by 1 MeV when $\kappa$ changes from 2 to 1.

In FAPT, which is an analytic QCD model with exceptionally fast convergence properties, the more simple approach (2/1-loop) gives the results close to (within a few per cent) the approaches using the higher-loop versions for the underlying pQCD. Therefore, in FAPT model, we can use various levels (2/1-, 3/2- and 4/3-loop), while for the other two versions of analytic QCD (2$\delta$anQCD and MPT) the preferred versions are 4/3-loop.

In addition, in FAPT, the program allows to choose either the usual version (i.e., at a fixed chosen $N_f$), or a ``global'' version \cite{Sh} for which the underlying pQCD coupling $a(Q^2;N_f)$ [and its discontinuity function $\rho_1^{(N_f,{\rm pt})}(\sigma) \equiv {\rm Im} a(-\sigma - i \epsilon;N_f)$] is replaced by a new, ``global'', pQCD coupling 
\bea
\lefteqn{
\!\!\!\!\!\!\!\!\!
a^{({\rm glob.})}(Q^2) = a(Q^2;N_f=3;{\overline \Lambda}_3) \Theta(|Q^2|\leq \mu^{(4)2})
+a(Q^2; N_f=4; {\overline \Lambda}_4) \Theta(\mu^{(4)2} < |Q^2| \leq \mu^{(5)2})
}
\nonumber\\
&&
+ a(Q^2;N_f=5;{\overline \Lambda}_5) \Theta(\mu^{(5)2} < |Q^2| \leq \mu^{(6)2})+
a(Q^2; N_f=6;{\overline \Lambda}_6) \Theta(\mu^{(6)2} < |Q^2|) \ ,
\label{ApertGlob}
\eea
where $\mu^{(3)}=\kappa m_3 = \kappa \m_c(\m_c)$, etc., 
and the scales ${\overline \Lambda}_{N_f}$ and the RGE-running of $a(Q^2; N_f)$ 
are determined by $N/(N-1)$-loop approach in $\MSbar$ 
(in the following referred to simply as $N$-loop approach; $N=1,2,3,4$). 
However, in such a global 
FAPT the values of the scales ${\overline \Lambda}_{N_f}$ differ somewhat 
from those of the actually valid pQCD [in the latter, the world average 
value $a(M_Z^2;5) = 0.1184/\pi$ in $\MSbar$ scheme fixes the scale 
${\overline \Lambda}_{5}$, see Table \ref{bLtab}]. 
The preferred values in the global FAPT are
${\overline {\Lambda}}_5 \approx 0.260$ GeV \cite{Sh,BMS06,BMS10,Bakulev}, 
corresponding to ${\overline {\Lambda}}_3 \approx 0.435$ GeV
[and $\alpha_s(M_Z^2;5;\MSbar) \approx 0.1218$] in
2/1-loop approach with $\kappa=2$,
and to  ${\overline {\Lambda}}_3 \approx 0.400$ GeV
in 4/3-loop approach with $\kappa=2$.

\section{Analytic QCD models}
 \label{sec:anQCD}

\subsection{General formalism}
\label{sec:genfor}

\begin{figure}[htb]
\centering\includegraphics[width=120mm]{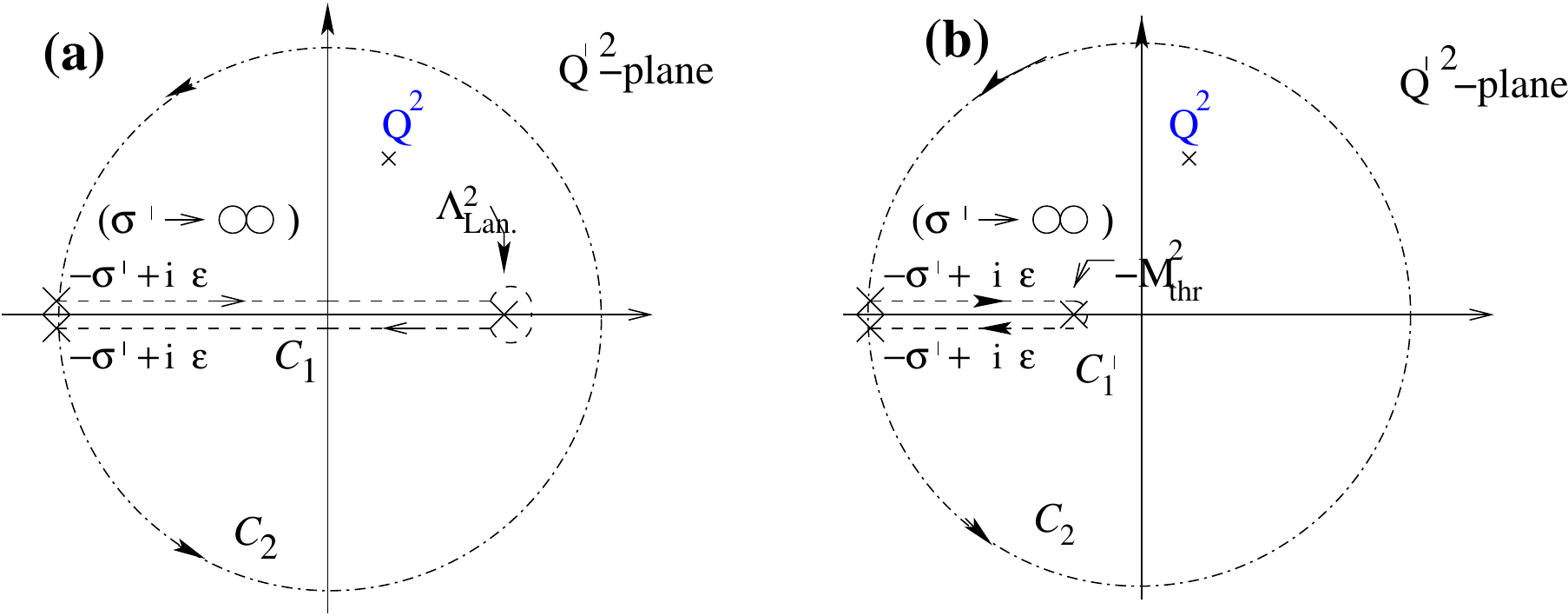}
\vspace{-0.4cm}
\caption{(a) The integration contour for the 
integrand $a(Q'^2)/(Q'^2 - Q^2)$ leading to the 
dispersion relation (\ref{adisp}) for  $a(Q^2)$; 
(b) the integration contour for the integrand  $\A(Q'^2)/(Q'^2 - Q^2)$
of a holomorphic coupling $\A(Q'^2)$
leading to the dispersion relation (\ref{Adisp}). The radius of the
circular section tends to infinity.}
\label{intpathab}
\end{figure}

In analytic QCD models, the dispersion relation between the
discontinuity function 
$\rho_1(\sigma) \equiv {\rm Im} \A_1(-\sigma - i \epsilon)$
and the coupling itself $\A_1(Q^2)$ plays usually a fundamental role,
where the discontinuity function $\rho_1(\sigma)$ is proportional
to the discontinuity of $\A_1$ across the cut at $Q^2=-\sigma$ ($<0$).
In pQCD such dispersion relation also exists. Namely,
when the function $a(Q^{'2})/(Q^{'2}-Q^2)$ is integrated 
in the $Q^{'2}$ complex plane along an appropriate closed contour 
which avoids all the cuts and encloses the pole $Q^{'2}=Q^2$
(cf.~Fig.~\ref{intpathab}(a)), and the Cauchy theorem is applied,
the following dispersion relation is obtained:
\begin{equation}
a(Q^2) = \frac{1}{\pi} \int_{\sigma= - {\Lambda^2_{\rm Lan.}} - \eta}^{\infty}
\frac{d \sigma {\rho_1^{\rm {(pt)}}}(\sigma) }{(\sigma + Q^2)},
   \quad (\eta \to +0).
\label{adisp}
\end{equation}
Here, $\rho_1^{\rm {(pt)}}(\sigma) \equiv {\rm Im} a(-\sigma - i \epsilon)$
is the discontinuity function of the pQCD coupling $a$
along the entire cut axis,
and $Q'^2=\Lambda^2_{\rm Lan.}$ ($>0$) is the branching point of the Landau 
cut of the pQCD coupling $a(Q^2)$.

In general analytic QCD models the dispersion relation has the form
\be
\A_1(Q^2) = \frac{1}{\pi} \int_{\sigma=M^2_{\rm thr}}^{\infty} \frac{d \sigma \rho_1(\sigma)}{(\sigma + Q^2)} \ ,
\qquad {\rm where: \ } \rho_1(\sigma) \equiv {\rm Im} \mathcal{A}_1(-\sigma - i \varepsilon) \ .
\label{Adisp}
\ee
The discontinuity function $\rho_1(\sigma)$ 
is defined for $\sigma \geq 0$; usually, the discontinuity cut is
nonzero below a threshold value $-\sigma \leq -M_{\rm thr}^2$ where
$M_{\rm thr} \sim M_{\pi}$. Therefore $Q^2$ can have any value in the 
complex plane except on the cut $(-\infty, -M^2_{\rm thr}]$
(cf.~Fig.~\ref{intpathab}(b)).

We regard either the discontinuity function $\rho_1(\sigma)$, or the 
coupling function $\A_1(Q^2)$, as the quantity which defines the anQCD model.
Below we describe how one constructs from them other quantities, such as
analytic analogs $\A_{\nu}(Q^2)$ of powers $a(Q^2)^{\nu}$ (where
$\nu$ is a real number) once the function $\rho_1(\sigma)$ or
$\A_1(Q^2)$ is known.

In order to find the correct analogs $\A_n$ of the powers $a^n$,
the logarithmic derivatives are needed
\be
\tilde{\mathcal{A}}_{n+1}(Q^2) \equiv \frac{(-1)^n}{\beta_0^n n!}
\left( \frac{ \partial}{\partial \ln Q^2} \right)^n
\A_1(Q^2) \ , \qquad (n=0,1,2,\ldots) \ .
\label{tAn}
\ee
We note that
for $n=0$ we have $\tilde{\mathcal{A}}_1 \equiv \mathcal{A}_1$.
We can write the
logarithmic derivatives in the following form~\cite{GCAK}:
\be
\tilde{\mathcal{A}}_{n+1}(Q^2) = \frac{1}{\pi} \frac{(-1)}{\beta_0^n \Gamma(n+1)}
\int_{0}^{\infty} \ \frac{d \sigma}{\sigma} \rho_1(\sigma)  
{\rm Li}_{-n} ( -\sigma/Q^2 ) \ .
\label{disptAn2}
\ee
This relation is valid for $n=0,1,2,...$. Analytic continuation
in $n \mapsto \nu$ ($\nu \in \Re$) gives us\footnote{
In \texttt{Mathematica} \cite{Math}, the ${\rm Li}_{-\nu}(z)$ function
is implemented as ${\rm PolyLog}[-\nu,z]$. 
However, in  \texttt{Mathematica} 9.0.1, at large $|z| > 10^7$,
${\rm PolyLog}[-\nu,z]$ is unstable. For such $z$ we should 
use the identities relating ${\rm Li}_{-\nu}(z)$ with ${\rm Li}_{-\nu}(1/z)$, 
which can be found, for example, in \cite{Erdelyi}. Our supplementary module
\texttt{Li\_\_nu.m} gives such stable functions 
${\rm Li}_{-\nu}(z)={\rm polylog}[-\nu,z]$. In \texttt{Mathematica} 10.0.1 this
problem is solved.} 
the logarithmic noninteger derivatives \cite{GCAK}
\be
\tilde{\mathcal{A}}_{\nu+1}(Q^2) = \frac{1}{\pi} \frac{(-1)}{\beta_0^{\nu} \Gamma(\nu+1)}
\int_{0}^{\infty} \ \frac{d \sigma}{\sigma} \rho_1(\sigma)  
{\rm Li}_{-\nu}\left( - \frac{\sigma}{Q^2} \right) \quad (-1 < \nu) \ .
\label{tAnu1}
\ee
We note that the integral converges for $\nu > -1$. Namely, at high
$\sigma$ ($|z| \gg 1$ where $z \equiv \sigma/Q^2$) we have in the integrand of
equation (\ref{tAnu1}): $\rho_1(\sigma) \approx
\rho_1^{\rm (pt)}(\sigma) \sim \ln^{-2} \sigma \sim \ln^{-2} z$ and 
${\rm Li}_{-\nu}(-z) \sim \ln^{-\nu} z$ (for noninteger $\nu$). 
Therefore, the integral converges at 
$\sigma \to \infty$ if $\nu > -1$. The integral obviously converges at low $\sigma$, too.\footnote{A related, but somewhat lengthier, formula for 
$\tilde{\mathcal{A}}_{\nu+1}(Q^2)$ in terms of $\rho_1(\sigma)$ 
which is valid in an extended interval $(-2 < \nu)$,
was also obtained in Ref.~\cite{GCAK} [cf.~Eq.(22) there]. Our
\texttt{Mathematica} package uses that lengthy formula.}

We can recast the result (\ref{tAnu1}) into an alternative form involving
the spacelike coupling $\mathcal{A}_1$ instead of the discontinuity function
$\rho_1(\sigma)$. This gives us
(for  $\nu = n + \delta$, with $0 < \delta < 1$ and 
$n=-1, 0, 1, 2, \ldots$) \cite{GCAK}
\bes
\label{tAnu2}
\bea
\tilde{\mathcal{A}}_{\nu+1}(Q^2) &\equiv& \tilde{\mathcal{A}}_{n+1+\delta}(Q^2) 
\nonumber\\
& = &
\frac{1}{\beta_0^{\nu} \Gamma(1+\nu)\Gamma(1-\delta)} 
\left(- \frac{d}{d \ln Q^2}\right)^{n+1}
\int_0^1 \frac{d \xi}{\xi} \mathcal{A}_1(Q^2/\xi) \ln^{-\delta}\left(\frac{1}{\xi}\right) 
\label{Anu2a}
\\
& = &
\frac{1}{\beta_0^{\nu}} \frac{\Gamma(1+\delta)}{\Gamma(n+1+\delta)}
\frac{ \sin(\pi \delta) }{(\pi \delta)}
\left(- \frac{d}{d \ln Q^2}\right)^{n+1}
\int_0^{\infty} \frac{dt}{t^{\delta}} \mathcal{A}_1(Q^2 e^t) \ ,
\label{Anu2b}
\eea
\ees
where the last form (\ref{Anu2b}) was obtained from the previous one by
the substitution $t = \ln(1/\xi)$ and using the identity 
$\Gamma(1 + \delta) \Gamma(1 - \delta) = \pi \delta/\sin(\pi \delta)$.

The analytic analogs  
$\mathcal{A}_\nu(Q^2) \equiv(a^\nu(Q^2))_{\rm an}$ of powers
$a(Q^2)^{\nu}$ can be constructed as linear combinations of 
$\tilde{\mathcal{A}}_{\nu+m}$'s:
\be
\A_\nu=\tA_{\nu}
+\sum_{m\geq1} {\tk}_m(\nu) \tA_{\nu+m},
\label{AnutAnu}
\ee
where the coefficients ${\tk}_m(\nu)$ were obtained in~\cite{GCAK} for general $\nu$.

Tha approach (\ref{tAnu1}) with (\ref{AnutAnu}) 
[$\Leftrightarrow$  (\ref{tAnu2}) with (\ref{AnutAnu})] for the case of
integer $\nu$ was constructed in Refs.~\cite{CV1,CV2}, and for general
real $\nu$ in Ref.~\cite{GCAK}. 

Specifically, let us consider a general
spacelike scale- and scheme-invariant physical quantity ${\cal D}(Q^2)$ 
which has the
available truncated perturbation (power) series of the form
\be
{\cal D}^{[N]}(Q^2;\kappa)_{\rm pt} = a(\kappa Q^2)^{\nu_0} + d_1(\kappa) a(\kappa Q^2)^{\nu_0+1} +
\ldots + d_{N-1}(\kappa)  a(\kappa Q^2)^{\nu_0+N-1} \ ,
\label{Dpt}
\ee
where $0<\kappa \sim 1$ is the renormalization scale parameter. The evaluation
of this quantity in a general analytic QCD model is then performed by the
substitution $a^{\nu_0+n} \mapsto \A_{\nu_0+n}$
\be
{\cal D}^{[N]}(Q^2;\kappa)_{\rm an} = \A_{\nu_0}(\kappa Q^2) + d_1(\kappa) \A_{\nu_0+1}(\kappa Q^2) +
\ldots + d_{N-1}(\kappa)  \A_{\nu_0+N-1} (\kappa Q^2) \ ,
\label{Dan}
\ee
with the quantities $\A_{\nu_0+n}$ constructed according to Eq.~(\ref{AnutAnu})
where the truncations are made, in general, at the highest available order
of the series (\ref{Dpt}), i.e., at $\sim a^{\nu_0+N-1} \sim \tA_{\nu_0+N-1}$
\be
\A_{\nu_0+n} = \tA_{\nu_0+n} + \sum_{m=1}^{N-1-n} \tk_m(\nu_0+n) \tA_{\nu_0+n+m} \ .
\label{AnutAnutr}
\ee
We refer for more details to Refs.~\cite{GCAK,Cvetic}. It is important to note
that $\A_{\nu_0+n} \not= (\A_1)^{\nu_0+n}$, i.e., the series (\ref{Dan}) is a 
nonpower series in any analytic QCD which is not perturbative. If, instead, 
we used in such analytic QCD the powers $(\A_1)^{\nu_0+n}$, 
the resulting truncated power series would
show increased renormalization scale dependence and 
(for low $|Q^2|$) strongly divergent behavior when $N$ increases, 
a consequence of incorrect
treatment of the nonperturbative constributions contained in the
difference $\A_1(\mu^2) - a(\mu^2)$, as emphasized in Refs.~\cite{Cvetic}. 

Further, the result (\ref{Dan})-(\ref{AnutAnutr}) can be reexpressed
in terms of $\tA_{\nu_0+n}$'s 
\be
{\cal D}^{[N]}(Q^2;\kappa)_{\rm an} = \tA_{\nu_0}(\kappa Q^2) + {\widetilde d}_1(\kappa) \tA_{\nu_0+1}(\kappa Q^2) +
\ldots + {\widetilde d}_{N-1}(\kappa)  \tA_{\nu_0+N-1} (\kappa Q^2) \ ,
\label{Dan2}
\ee
where
\be
{\widetilde d}_M(\kappa) = d_M(\kappa) + 
\sum_{q=1}^{M} \tk_q(\nu_0+M-q) d_{M-q}(\kappa) \ , \qquad (M=1,2,\ldots,N-1) \ ,
\label{td}
\ee
and the convention $d_0(\kappa)=1$ is taken. Comparing the expressions
(\ref{Dan}) and (\ref{Dan2}), it becomes clear that in anQCD the basic
quantities in perturbation 
expansion are the (generalized) logarithmic derivatives
$\tA_{\nu}$, and not the (nonpower) analogs $\A_{\nu}$ of pQCD powers $a^{\nu}$.
 These aspects have been presented and emphasized in more detail in
Refs.~\cite{Cvetic}.

When we evaluate a timelike physical quantity ${\cal F}(\sigma)$, such a
quantity can be expressed as a contour integral of the corresponding
spacelike quantity ${\cal D}(Q^2)$ in the complex $Q^2$ plane.
Therefore,  ${\cal F}(\sigma)$ can be expressed as a series of 
contour integrals of the couplings $\A_{\nu}(Q^2)$ or $\tA_{\nu}(Q^2)$.
\subsection{Fractional Analytic Perturbation Theory (FAPT)}
 \label{sec:FAPT}

The APT procedure \cite{ShS} is the elimination
of the contributions of the Landau cut 
$0 < (-\sigma) \leq \Lambda_{\rm Lan.}^2$.
This gives the APT analytic analog $\mathcal{A}_1^{\rm (APT)}(Q^2;N_f)$ of 
$a(Q^2;N_f)$ 
\be
\mathcal{A}_1^{\rm (APT)}(Q^2;N_f) = \frac{1}{\pi} \int_{\sigma= 0}^{\infty}
\frac{d \sigma {\rho_1^{\rm {(pt)}}}(\sigma;N_f) }{(\sigma + Q^2)} \ .
\label{A1APT}
\ee
This procedure can be extended to the construction of
the APT-analogs $\mathcal{A}_n^{\rm (APT)}(Q^2)$ of $n$-integer powers
$a(Q^2)^n$ \cite{MS96,Sh} and their combinations (see also \cite{KS}).
The APT analogs of general powers
$a^{\nu}$ ($\nu$ a real exponent) are known as Fractional APT 
(FAPT)~\cite{BMS05,BKS05,BMS06,BMS10}; following the same procedure, they are
\begin{equation}
{\mathcal{A}}^{\rm {(FAPT)}}_{\nu}(Q^2;N_f) = \frac{1}{\pi} \int_{\sigma= 0}^{\infty}
\frac{d \sigma {\rho^{\rm {(pt)}}_{\nu}}(\sigma;N_f) }{(\sigma + Q^2)} \ ,
\label{AnuFAPT}
\end{equation}
where
\be
{\rho^{\rm {(pt)}}_{\nu}}(\sigma;N_f) = 
{\rm Im} \; a(Q^{'2}=-\sigma - i \epsilon; N_f)^{\nu} \ .
\label{rhonu1}
\ee
It turns out that in FAPT, 
where the approach (\ref{AnuFAPT}) can be applied,\footnote{
We note that in anQCD models other than FAPT as defined by Eq.~(\ref{A1APT}),
the approach of the type (\ref{AnuFAPT}) to the calculation of $\A_{\nu}$'s
is not applicable. This is so because in such anQCD models
$\rho_1(\sigma) \equiv {\rm Im} \A_1(-\sigma - i \epsilon)$
[$ \not= {\rm Im} a(-\sigma - i \epsilon)$] and, for $\nu \not=1$ we have: 
$\rho_{\nu}(\sigma) \equiv {\rm Im} \A_{\nu}(-\sigma - i \epsilon)$.
Therefore, $\rho_{\nu}(\sigma) \not=  {\rm Im} a(-\sigma - i \epsilon)^{\nu}$
and $\rho_{\nu}(\sigma) \not=  {\rm Im} \A_1(-\sigma - i \epsilon)^{\nu}$. 
The former inequality holds because the model is not FAPT; the latter
inequality holds because $\A_{\nu} \not= \A_1^{\nu}$ (for $\nu \not= 1$)
in general anQCD models which are simultaneously not pQCD. For models
which are anQCD and simultaneously pQCD (i.e., anpQCD), 
we refer to  Refs.~\cite{anpQCD}.} 
it is equivalent with the approach of Eqs.~(\ref{tAnu1}) and (\ref{AnutAnu})
[or, equivalently, Eqs.~(\ref{tAnu2}) and (\ref{AnutAnu})] that can be applied
in general anQCD models, if in the sums on the right-hand side of 
Eq.~(\ref{AnutAnu}) we do not make truncations of the type of 
Eq.~(\ref{AnutAnutr}), but rather include as many terms as possible.
We refer to Refs.~\cite{CV1,CV2,GCAK} for more details on these points.

In the global version of FAPT, the coupling 
${\mathcal{A}}^{\rm {(FAPT)glob.}}_{\nu}(Q^2)$
is obtained by applying the dispersion relation to the
discontinuity function of the power $\nu$ of the
global coupling (\ref{ApertGlob}), for $\sigma \geq 0$
\bea
{\rho^{\rm {(pt)}glob.}_{\nu}}(\sigma) &=& 
{\rm Im} \; a^{({\rm glob.})}(Q^2=-\sigma - i \epsilon)^\nu 
\nonumber\\
&\equiv& {\rho^{\rm {(pt)}}_{\nu}}(\sigma; {\overline \Lambda}_3)\Theta(|Q^2|\leq \mu^{(4)2}) +
 {\rho^{\rm {(pt)}}_{\nu}}(\sigma; {\overline \Lambda_4})\Theta(\mu^{(4)2}\leq |Q^2|\leq \mu^{(5)2})+
 \nonumber\\
 && {\rho^{\rm {(pt)}}_{\nu}}(\sigma; {\overline \Lambda_5})\Theta(\mu^{(5)2}\leq |Q^2|\leq \mu^{(6)2})+
   {\rho^{\rm {(pt)}}_{\nu}}(\sigma; {\overline \Lambda_6})\Theta(\mu^{(6)2}\leq |Q^2|) \ .
\label{rhonu2}
\eea
If the underlying pQCD running coupling $a(Q^2)$ runs according to the one-loop
perturbative RGE, the corresponding explicit expressions 
for $\mathcal{A}_{\nu}^{\rm (FAPT)}$ exist and were obtained
and used in Ref.~\cite{BMS05}
\be
\mathcal{A}_{\nu}(Q^2)^{\rm (FAPT, 1-\ell.)} = \frac{1}{\beta_0^{\nu}}
\left(  \frac{1}{\ln^{\nu}(z)} -
\frac{ {\rm Li}_{-\nu+1}(1/z)}{\Gamma(\nu)} \right) \ .
\label{MAAnu1l}
\ee
Here, $z \equiv Q^2/\Lambda^2$ and 
${\rm Li}_{-\nu+1}(x)$ is the polylogarithm function of order $-\nu+1$.
Explicit extensions to approximate higher loops were performed 
by expanding the one-loop result in a series of derivatives
with respect to the index $\nu$ \cite{BMS05,BMS06,BMS10}\footnote{
For practical purposes, we use in the integral
(\ref{AnuFAPT}) the $N$-loop level $\rho_{\nu}^{\rm (pt)}(\sigma)$ 
(where: $N \leq 4$).}
We refer for reviews of FAPT to Refs.~\cite{Bakulev,BaSh,Ste}. 

When in FAPT the underlying pQCD coupling $a(Q^2)$ is given by 
Eqs.~(\ref{1LpQCD}) and (\ref{a2lLambert}), the resulting theory
is called 1-loop and 2-loop FAPT, respectively. When $a(Q^2)$ is given
by Eq.~(\ref{aptexact}) with $c_2={\overline c}_2(N_f)$ of $\MSbar$ scheme, 
the resulting theory is called, somewhat loosely, 3-loop FAPT. When
$a(Q^2)$ is given by the expansion (\ref{NLpQCDapp}), 
with $c_2={\overline c}_2(N_f)$ and $c_3={\overline c}_3(N_f)$
($c_j=0$ for $j \geq 4$; and the truncation index ${\cal N}=8$ is used),
the resulting FAPT is called 4-loop.

Due to easiness of numerical implementation, in this model we incorporate
the FAPT-analytization of logarithmic powers, too
\begin{equation}
{\A}^{\rm {(FAPT)}}_{\nu,k}(Q^2) = \frac{1}{\pi} \int_{\sigma= 0}^{\infty}
\frac{d \sigma {\rm Im} \ \left[ a(-\sigma-i \epsilon)^{\nu} \ln^k  a(-\sigma-i \epsilon) \right] }{(\sigma + Q^2)} \ ,
\label{AnukFAPT}
\end{equation}
where $\nu$ is a general (noninteger) index and $k=0,1,2,\ldots$.

The couplings of ${\rm FAPT}_{N_f}$ and of global FAPT are calculated
also in the \texttt{Mathematica}  program of Ref.~\cite{BK2}. 
The values of couplings $\A_{\nu}(Q^2)$ 
of ${\rm FAPT}_{N_f}$ models in our program, 
when $\kappa=2$ is changed there to $\kappa=1$,
practically coincide with the corresponding values of \cite{BK2}.
In global FAPT,\footnote{
We note that our $\A_{\nu}$ corresponds to their $\A_{\nu}/\pi^{\nu}$;
and what we call (approximate) 3-loop (``3l'') they call 
more rigorously 3-loop-Pad\'e (``3P'').}
there are small differences between
our values and theirs, which tend to increase somewhat when $\nu$
increases: for $\nu<1$ the differences are $1 \%$ or less, for 
$1<\nu<2$ are $1$-$2 \%$, for $2 < \nu < 3$ are $2$-$3 \%$,
for $3 < \nu < 4$ are $4$-$8 \%$. We note, however, that with
increasing $\nu$ the couplings in FAPT decrease very fast. We believe that
one of the principal reasons for
the small mentioned differences lies in the fact that
in our program the quark thresholds (with $\kappa=1$) are implemented at
the masses $\m_q$ while in the program of Ref.~\cite{BK2} at the
quark pole masses.

Furthermore, the couplings of ${\rm (F)APT}_{N_f}$ are calculated also
by the programs of Ref.~\cite{BK1}, in Maple and in Fortran, and their
values practically coincide with ours.


\subsection{Two-delta analytic model (2$\delta$anQCD)}
 \label{sec:2danQCD}

\subsubsection{2$\delta$anQCD in low momentum regime ($N_f=3$)}
\label{2danQCDNf3}

In this anQCD model \cite{2danQCD}, the discontinuity function 
$\rho_1(\sigma) \equiv {\rm Im} \  
\mathcal{A}_1(Q^2=-\sigma - i \epsilon)$ (for $\sigma > 0$)
 agrees with the perturbative counterpart
$\rho_1^{\rm (pt)}(\sigma) \equiv {\rm Im} \  a(Q^2=-\sigma - i \epsilon)$
at sufficiently high scales 
$\sigma \geq M_0^2$ ($M_0^2 \sim 1 \ {\rm GeV}^2$);
while in the low-scale regime $0 < \sigma < M_0^2$ its otherwise
unknown behavior is parametrized as a linear combination of
(two) delta functions (a parametrization motivated by the Pad\'e
approximation approach for the running coupling \cite{Padappr})
\bes
\label{rho12d}
\bea
\rho_1^{(2\delta)}(\sigma; c_2) &=&
\pi \sum_{j=1}^2 f_j^2 \Lambda^2 \; 
\delta(\sigma - M_j^2) +  \Theta(\sigma-M_0^2) \times 
\rho_1^{\rm (pt)}(\sigma; c_2) 
\label{rho12da}
\\
& = & \pi \sum_{j=1}^2 f_j^2 \; \delta(s - s_j) +  
\Theta(s-s_0) \times r_1^{\rm (pt)}(s; c_2) \ ,
\label{rho2db}
\eea
\ees
where we define the dimensionless quantities:  $s=\sigma/\Lambda^2$, $s_j = M_j^2/\Lambda^2$ ($j=0,1,2$),
and $r_1^{\rm (pt)}(s; c_2) =  \rho_1^{\rm (pt)}(\sigma; c_2)
= {\rm Im} \ a(Q^2=-\sigma - i \epsilon; c_2)$. Here, 
$\Lambda^2$ ($\lesssim 10^{-1} \ {\rm GeV}^2$) is the Lambert scale
appearing in the expression ~(\ref{aptexact}) for $a$ 
[cf.~also Eq.~(\ref{zina2l})].
The underlying pQCD coupling is taken in the form (\ref{aptexact}) 
where the scheme parameter $c_2$ ($\equiv\beta_2/\beta_0$) is nonzero in 
general [cf. Eqs.~(\ref{betaPad})].

\begin{figure}[htb] 
\begin{minipage}[b]{.49\linewidth}
\centering\includegraphics[width=80mm]{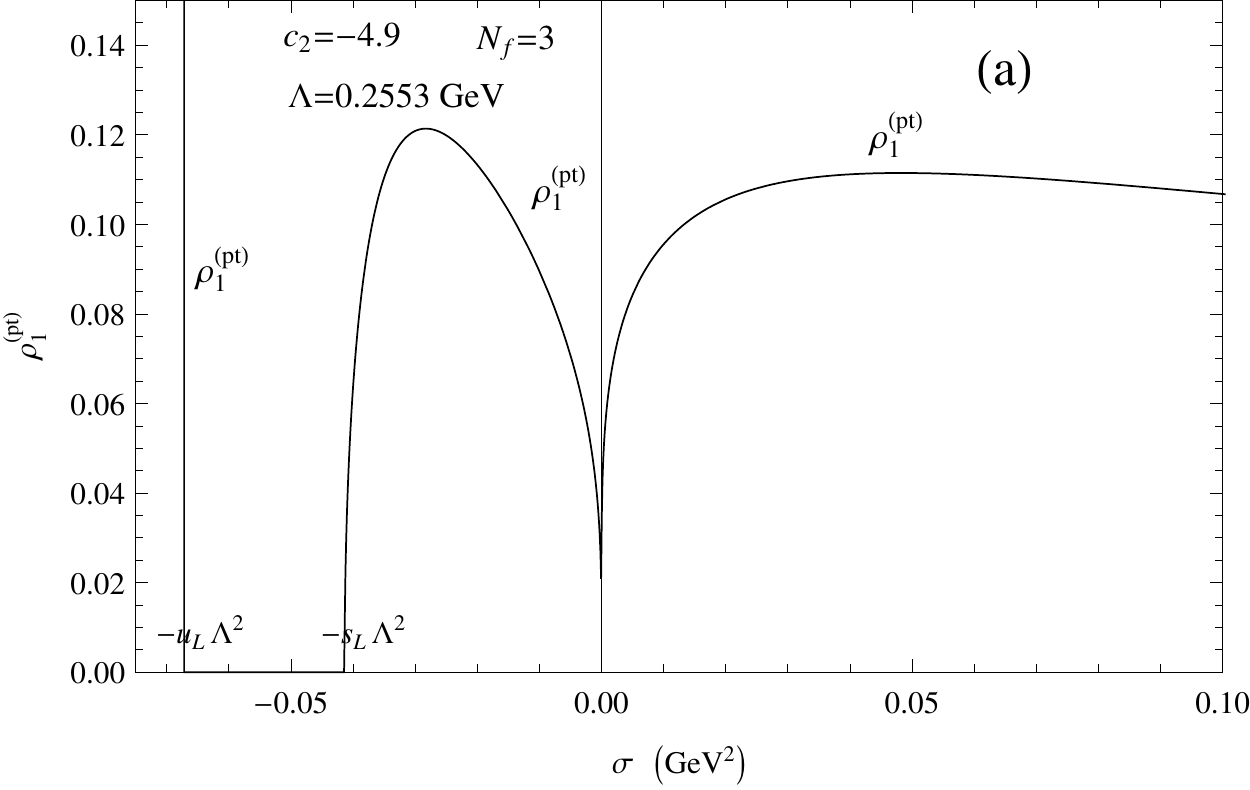}
\end{minipage}
\begin{minipage}[b]{.49\linewidth}
\centering\includegraphics[width=80mm]{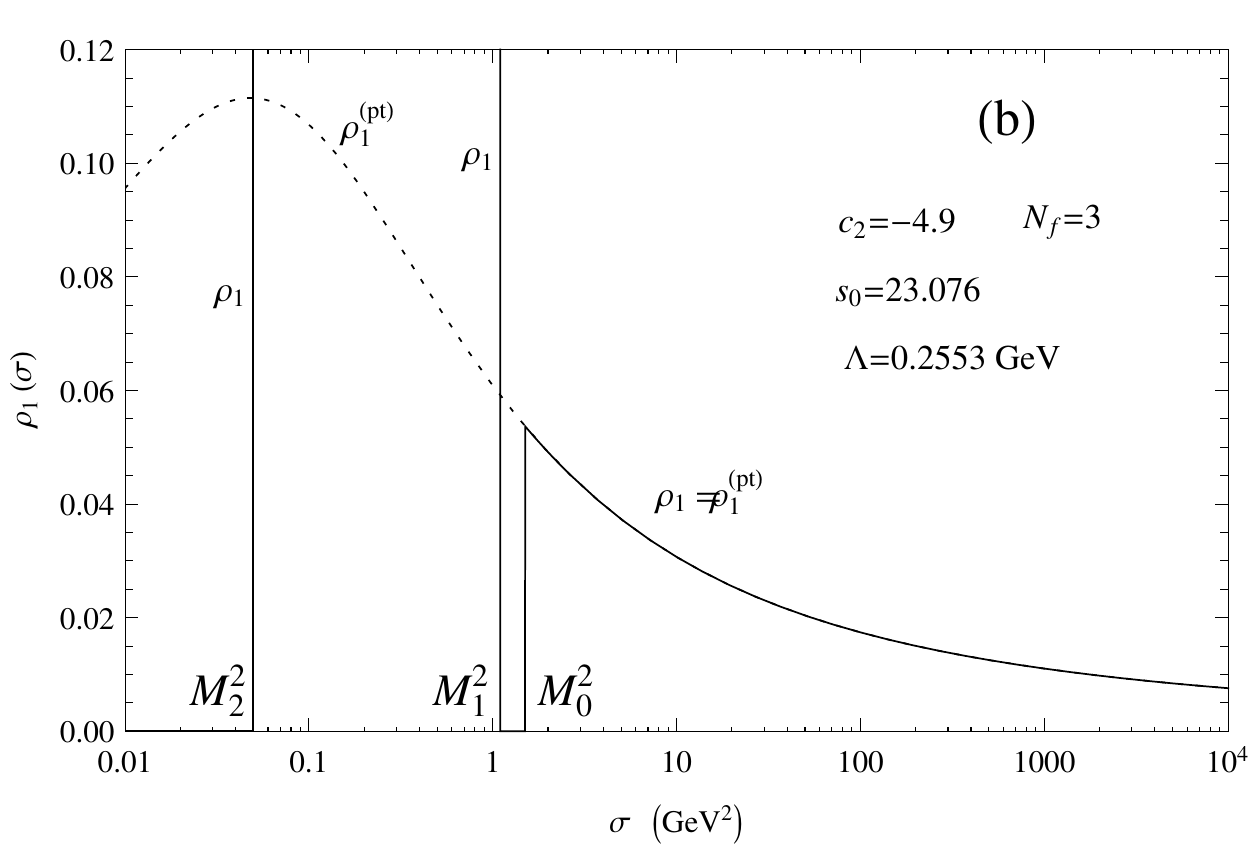}
\end{minipage}
\vspace{-0.4cm}
 \caption{\footnotesize (a) The discontinuity function 
$\rho_1^{\rm (pt)}(\sigma) \equiv {\rm Im} a(-\sigma - i \epsilon)$
of the perturbative coupling $a$ of Eq.~(\ref{aptexact}), for
$c_2=-4.9$ and $N_f=3$;
(b) the corresponding 2$\delta$anQCD discontinuity function
$\rho_1^{(2 \delta)}(\sigma)$ , Eq.~(\ref{rho12da}).
The MPT discontinuity function is
$\rho_1^{\rm (MPT)}(\sigma) = \rho_1^{\rm (pt)}(\sigma-m^2_{\rm gl})$,
cf.~Eq.~(\ref{MPT1}); when $m^2_{\rm gl}=0.7 \ {\rm GeV}^2$, 
this is just the curve of Fig.~(a) shifted by 
$0.7 \ {\rm GeV}^2$ toward the right.}
\label{figrho12d}
 \end{figure}

The aforementioned branching point of nonanalyticity 
$z=-1/e$ corresponds, according to Eq.~(\ref{zina2l}), 
to the scale $Q^2 = \Lambda^2 s_{\rm L}$ with
$s_{\rm L}=c_1^{-c_1/\beta_0}$ ($=0.6347$ when $N_f=3$). The
interval of Landau singularities of $a(Q^2)$  of Eq.~(\ref{aptexact})
is: $0< Q^2 < \Lambda^2 s_{\rm L}$.
In our case we will choose $c_2$ to be negative. In such a case
there is an additional pole-type
Landau singularity, at a somewhat higher scale $Q^2 =\Lambda^2 u_{\rm L}$ 
-- there the denominator in Eq.~(\ref{aptexact}) becomes zero,
cf.~Fig.~\ref{figACC1}(b).
Our preferred choice of the scheme in the model will be $c_2=-4.9$;
in this case we have $u_{\rm L} = 1.0311$ ($> s_{\rm L}$). 
For this ``canonical'' case, 
the underlying pQCD discontinuity function $\rho_1^{\rm (pt)}(\sigma)$ 
is presented in Fig.~\ref{figrho12d}(a) as a function of
$\sigma$,
and the corresponding 2$\delta$anQCD discontinuity function
$\rho_1^{(2 \delta)}(\sigma)$ in Fig.~\ref{figrho12d}(b). The
Lambert $\Lambda$ scale, appearing in Eq.~(\ref{zina2l}),
was taken with the value of $\Lambda = 0.255$ GeV because this
then corresponds to the world average value $a(M_Z^2;\MSbar)=0.1184/\pi$,
as will be seen later.

\begin{figure}[htb]
\centering\includegraphics[width=80mm]{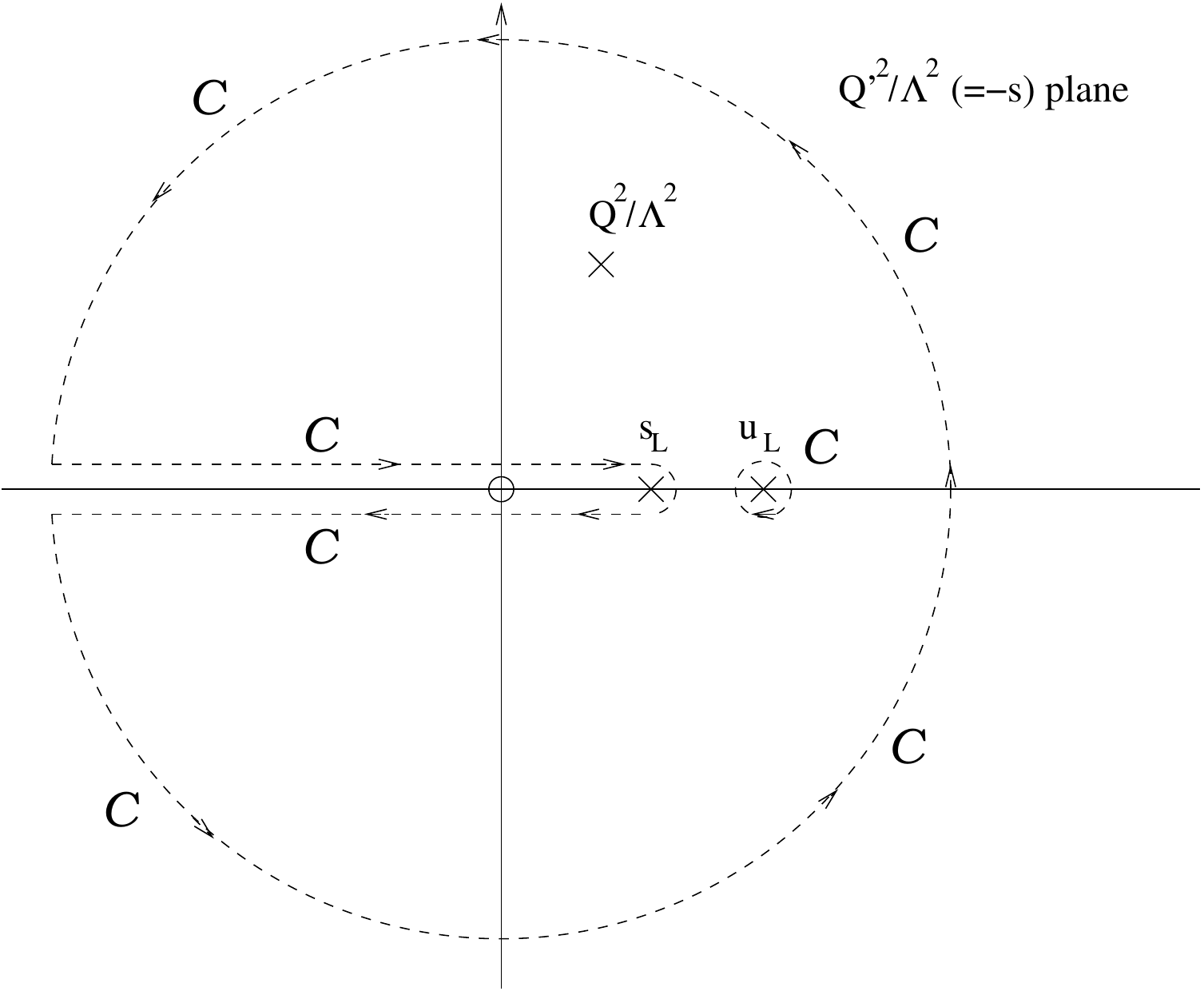}
\vspace{-0.4cm}
\caption{The integration contour for the 
integrand $a(Q'^2)/(Q'^2 - Q^2)$ leading to the 
dispersion relation (\ref{adisp2d}) for  $a(Q^2)$ of Eq.~(\ref{aptexact}) 
with $c_2 < 0$.
The radius of the large circular section tends to infinity.}
\label{intpath2d}
\end{figure}

In Fig.~\ref{figrho12d}(a) we see
that $a(Q^2;N_f)$, for $c_2=-4.9$, has a Landau pole at 
$\sigma (\equiv - Q^2) = - u_{\rm L} \Lambda^2$ ($\approx -0.067 \ {\rm GeV}^2$)
 and the Landau branching point at $\sigma = -s_{\rm L} \Lambda^2$
($\approx -0.041 \ {\rm GeV}^2$).
Therefore,
the dispersive relation (\ref{adisp}) for the underlying perturbative
coupling $a(Q^2;N_f=3)$ obtains a slightly generalized form [in comparison
with Eq.~(\ref{adisp})]
\bea
\label{adisp2d}
a(Q^2) &=& 
\frac{1}{\pi} 
\int_{s= - s_{\rm L} - \eta}^{\infty} ds \;
\frac{r_1^{\rm (pt)}(s;c_2)}{(s + Q^2/\Lambda^2)} + 
\frac{{\rm Res}_{(z=u_{\rm L})} a(z \Lambda^2; c_2)}{ (- u_{\rm L} + Q^2/\Lambda^2) } \ ,
\eea
which is obtained by application of the Cauchy theorem to the function
 $a(Q'^2)/(Q'^2 - Q^2)$ along the contour depicted in Fig.~\ref{intpath2d}
[in contrast to the simple contour Fig.~\ref{intpathab}(a) leading to
Eq.~(\ref{adisp})].

The perturbative discontinuity function 
$r_1^{\rm (pt)}(s;c_2) = {\rm Im} \ a(Q^2=-s \Lambda^2 - i \epsilon; c_2)$,
which is nonzero for $-s_{\rm L}  < s < + \infty$ and
at $s=-u_{\rm L}$, has the specific form
\be
r_1^{\rm (pt)}(s;c_2) = 
{\Bigg \{} 
\begin{array}{ll} 
{\rm Im} \left[ \frac{(-1)}{c_1} 
\frac{1}{\left[ 1 - (c_2/c_1^2) + W_{+1} \left( \frac{-1}{c_1 e} 
|s|^{-\beta_0/c_1} - i \epsilon \right) \right]} \right] 
& (s < 0) \ , 
\\
{\rm Im} \left[ \frac{(-1)}{c_1} 
\frac{1}{\left[ 1 - (c_2/c_1^2) + W_{+1} \left( \frac{-1}{c_1 e} |s|^{-\beta_0/c_1} \exp( i \beta_0 \pi/c_1) \right) \right]} \right] & (s > 0)\ .
\end{array}
\label{r1s}
\ee
The analytic (spacelike) coupling $\mathcal{A}_1^{(2\delta)}(Q^2;c_2)$ of
the two-delta anQCD model is constructed  on
the basis of the discontinuity function (\ref{rho12d})
[cf.~Eq.~(\ref{r1s}) for $s>0$] using the dispersion relation.
This gives
\be
\mathcal{A}_1^{(2\delta)}(Q^2;c_2) = \sum_{j=1}^2 \frac{f_j^2}{(s_j+u)} +
\frac{1}{\pi} \int_{s_0}^{\infty} ds \; 
\frac{r_1^{\rm (pt)}(s;c_2)}{(s+u)} \ ,
\label{A12d}
\ee
where $u = Q^2/\Lambda^2$. 

In the Two-delta $N_f=3$ anQCD model with a chosen value of $c_2$
[$2\delta{\rm anQCD}_{N_f=3}(c_2)$],
and with $c_1=c_1(N_f=3)=(\beta_1/\beta_0)_{N_f=3}$, 
the first three quark flavors are approximated as massless. 
Most importantly, the model is constructed so that at high
$|Q^2|$ it basically coincides with the underlying ${\rm pQCD}_{N_f=3}(c_2)$,
and that it simultaneously reproduces the experimental value of
the (canonical) decay ratio $r_{\tau}$ of the
strangeless and massless $(V+A)$-channel semihadronic decays
of the $\tau$ lepton: $r_{\tau} = 0.203$.
This is achieved in three steps. 
\begin{enumerate}
\item
The first step is to obtain the value of the Lambert scale
 $\Lambda$ appearing in the underlying ${\rm pQCD}_{N_f=3}(c_2)$
coupling $a(Q^2)$ of Eqs.~(\ref{aptexact}) and (\ref{zina2l}).
This is done in the following way: the world average
value ${\overline a}(M_Z^2)=0.1184/\pi$ is evolved by 4-loop $\MSbar$ RGE 
from $Q^2=M_Z^2$ down to $Q^2=(2 \m_c^2)$, obtaining
${\overline a}_{\rm in} \equiv {\overline a}((2 \m_c)^2;N_f=3) 
= 0.26535/\pi$.
3-loop threshold matching (\ref{msb}) is used, 
at $Q^2=(2 \m_b)^2$ and $(2 \m_c)^2$
($\m_b=4.2$ GeV and $\m_c=1.27$ GeV).
From this value  ${\overline a}_{\rm in}$, in $\MSbar$ scheme,
the corresponding value
$a_{\rm in} \equiv a((2 m_c)^2;c_2, c_2^2/c_1, \ldots;N_f=3)$ in the
renormalization scheme of the $2\delta{\rm anQCD}_{N_f=3}(c_2)$ model
is obtained, i.e., in the scheme determined by the beta function
$\beta(a)$ of Eq.~(\ref{betaPad}). 
This is performed by solving for $a_{\rm in}$ the integrated form of RGE 
(i.e., implicit solution) in its subtracted
form, cf.~Appendix A of Ref.~\cite{Stevenson}
(cf.~also Appendix A of Ref.~\cite{CK})
\bea
\lefteqn{
\frac{1}{a_{\rm in}} + c_1 \ln \left( \frac{c_1 a_{\rm in}}{1\!+\!c_1 a_{\rm in}} \right)
+ \int_0^{a_{\rm in}} dx \left[ \frac{ \beta(x) + \beta_0 x^2 (1\!+\!c_1 x)}
{x^2 (1\!+\!c_1 x) \beta(x) } \right] =}
\nonumber\\
&& \frac{1}{{\overline a}_{\rm in}} + c_1 \ln \left( \frac{c_1 {\overline a}_{\rm in}}{1\!+\!c_1 {\overline a}_{\rm in}} \right)
+ \int_0^{{\overline a}_{\rm in}} dx \left[ \frac{ {\overline \beta}(x) + \beta_0 x^2 (1\!+\!c_1 x)}
{x^2 (1\!+\!c_1 x) {\overline \beta}(x) } \right].
\label{match}
\eea
For $c_2=-4.9$ this gives $a_{\rm in} =0.24860/\pi$. 
Equating this value with the expression
(\ref{aptexact}) (with $c_2=-4.9$ and $N_f=3$) gives the Lambert
scale $\Lambda \equiv \Lambda_3$ of the model:
$\Lambda=0.2553$ GeV. For other values of $c_2$, other values
of $\Lambda$ are obtained.
\item
The second step is to make the model $2\delta{\rm anQCD}_{N_f=3}(c_2)$ 
practically coincide with the underlying ${\rm pQCD}_{N_f=3}(c_2)$
at high $|Q^2| > \Lambda^2$.
In general, $\mathcal{A}_1(Q^2;c_2)$ differs from
$a(Q^2;c_2)$ at $Q^2 > \Lambda^2$ by $\sim (\Lambda^2/Q^2)^1$,
as is the case, e.g., with FAPT and MPT.
In 2$\delta$anQCD we impose the condition 
\be
\mathcal{A}_1(Q^2;c_2) - a(Q^2;c_2) \sim 
(\Lambda^2/Q^2)^{n_{\rm max}} \quad {\rm with} \; n_{\rm max}=5 \ .
\label{ITEPc}
\ee
The condition (\ref{ITEPc}) represents in practice four conditions,
which fix four dimensionless parameters $s_j, f_j^2$ ($j=1, 2$) in terms of the
fifth dimensionless parameter $s_0$.
\item
The third step is to ensure that the model $2\delta{\rm anQCD}_{N_f=3}(c_2)$ 
reproduces the correct central value of the 
$(V+A)$-channel semihadronic $\tau$ decay
ratio\footnote{This quantity is normalized canonically, i.e., its
perturbation expansion is $(r_{\tau})_{\rm pt} = a + {\cal O}(a^2)$.
For details on $r_{\tau}$ and its evaluation in analytic QCD approaches,
we refer to Ref.~\cite{2danQCD} and Appendices B-E of Ref.~\cite{anpQCD}.}
$r_{\tau}(\Delta S=0, m_q=0)_{exp}=0.203\pm 0.004$.
\end{enumerate}  

The scheme parameter $c_2$ ($\equiv \beta_2/\beta_0$) can still be varied. 
Physical considerations guide us to restrict the preferred values
of the pQCD-onset scale $M_0$ and of the coupling $\A_1(Q^2)$ at $Q^2=0$: 
$M_0\leq 1.5$ GeV and $\A_1(0) < 1$.
This gives us the variation of $c_2$ in the interval $-5.6<c_2<-2,0$. 
In Table~\ref{2dtab} we present the results for the parameters of the model 
for various values of $c_2$ in this interval.\footnote{
In Ref.~\cite{2danQCD}, the obtained parameters of the model were slightly different. 
The principal reason for that was that the 3-loop quark threshold conditions in the 
$\MSbar$ RGE-running downwards
in Ref.~\cite{2danQCD} were implemented by a version of (\ref{msb}) expressing $a$
as a truncated power series of $a^{'}$. However, the numerical results for the coupling,
at a given $c_2$, are almost indistinguishable from those of Ref.~\cite{2danQCD}.}   
Our preferred choice is $c_2=-4.9$ where $M_0 \approx 1.23$ GeV and
$\A_1(0) \approx 0.82$.

The (generalized) logarithmic derivatives $\tA_{\nu}$ are then constructed by 
the procedure (\ref{tAnu1}), and the power analogs $\A_{\nu}$ by the linear
combinations (\ref{AnutAnutr}) (where $\nu_0= \nu$)
with the truncation (``loop'') index
there being $N=1,2,3,4,5$.

\begin{table}
\caption{Values of the parameters of the considered 2$\delta$anQCD model,
for $N_f=3$ and $-5.6 \leq c_2 \leq -2.0$.
We consider $c_2=-4.9$ ($M_0 \approx 1.23$ GeV) 
as the preferred representative case. 
The value $\pi \times a_{\rm in} = \alpha_s((2 m_c)^2;c_2,\ldots; N_f=3)$ 
and the Lambert scale
value $\Lambda$ in the corresponding cases are for
the QCD coupling parameter value $\alpha_s^{({\overline {\rm MS}})}(M_Z^2) = 0.1184$.}
\label{2dtab}  
\begin{center}
\centering
\begin{tabular}{|c|c|c|ccccc|cc|}
\hline
$c_2$   & $\pi \times a_{\rm in}$  & $\Lambda$ [GeV] & $s_0$ & $s_1$ & $f_1^2$ & $s_2$ & $f_2^2$  & $M_0$ & $\mathcal{A}_1(0)$
\\ 
\hline
-5.60 & 0.2477 & 0.2339 & 24.416 & 17.787 & 0.3013 & 0.6906 & 0.6150 & 1.156 & 0.9999
\\
-5.40 & 0.2480 & 0.2398 & 24.054 & 17.533 & 0.2936 & 0.7179 & 0.5960 & 1.176 & 0.9389
\\
-4.90 & 0.2486 & 0.2552 & 23.076 & 16.839 & 0.2746 & 0.7688 & 0.5505  & 1.226 & 0.8231
\\
-4.00 & 0.2498 & 0.2857 & 21.142 & 15.454 & 0.2416 & 0.8094 & 0.4753  & 1.314 & 0.6916
\\
-3.00 & 0.2512 & 0.3237 & 18.903 & 13.836 & 0.2078 & 0.8003 & 0.4020  & 1.407 & 0.6042
\\
-2.00 & 0.2526 & 0.3668 & 16.708 & 12.241 & 0.1775 & 0.7557 & 0.3388 & 1.499 &0.5481 
\\ 
\hline
\end{tabular}
\end{center}
\end{table}
\subsubsection{2$\delta$anQCD for $N_f \geq 4$}
 \label{2danQCDNf4}

\begin{figure}[htb]
\centering\includegraphics[width=140mm]{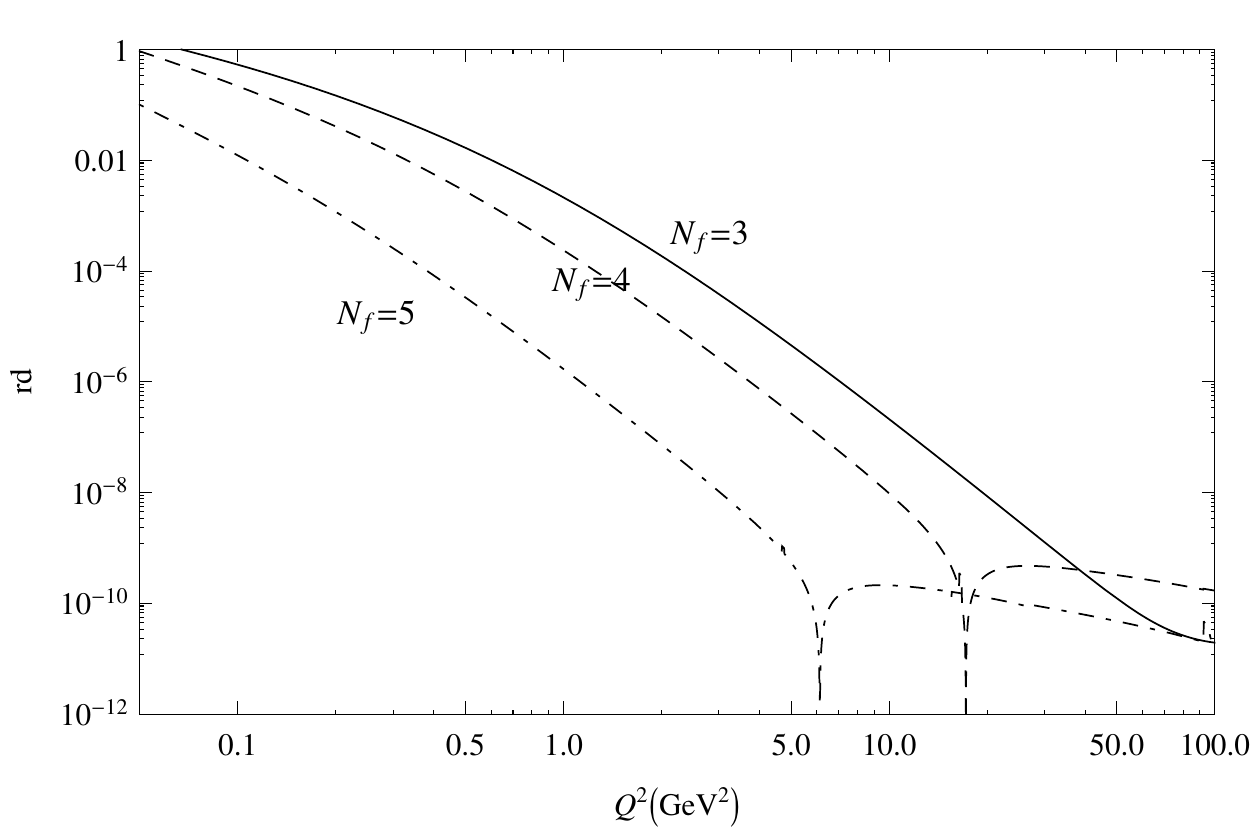}
\vspace{-0.4cm}
\caption{The relative difference between 2$\delta$anQCD coupling and the
underlying pQCD coupling, ${\rm rd}(Q^2) \equiv |\A_1(Q^2;N_f)/a(Q^2;N_f) - 1|$,
as a function of positive $Q^2$, for $N_f=3,4,5$. The parameter $c_2$ of the model
is set equal to $c_2(N_f)=-4.9$.}
\label{reldiff}
\end{figure}

The 2$\delta$anQCD model can be constructed also for $N_f=4,5,6$. In such cases,
for a chosen value of $c_2$ [$=c_2(N_f)$], the value of $\Lambda_{N_f}$ is 
determined by pQCD, as in $N_f=3$ case. Further, the condition (\ref{ITEPc}) 
again gives us the values of the four parameters $s_j$ and $f_j^2$ ($j=1,2$) 
in terms of $s_0$. However, since in the case of $N_f \geq 4$ the couplings 
$\A_{\nu}(Q^2)$ should be applied only for $|Q^2| > (2 m_{N_f})^2$ 
(where: $m_4 = \m_c$, $m_5 = \m_b$, $m_6=\m_t$), the
low-momentum quantity $r_{\tau}$ cannot and should not be evaluated in such
framework. Therefore, for $N_f \geq 4$ the value of the $s_0$ parameter is
free. In our program, we kept the value of $s_0(N_f)$ equal to the
corresponding value of $s_0(N_f=3)$. In such cases, the $N_f=4$ 2$\delta$anQCD
model still remains formally analytic, while for $N_f=5,6$ it is formally
nonanalytic (since $s_2<0$ is such a case). Nevertheless, we prefer to
keep such, relatively low, values of $s_0$ for $N_f \geq 4$, because then
the coefficient on the right-hand side of Eq.~(\ref{ITEPc}) in front of
$(\Lambda^2/Q^2)^{5}$ is not very large; therefore, the model for
$N_f \geq 4$ practically agrees with the underlying pQCD. The relative
difference between 2$\delta$anQCD values $\A_1(Q^2;N_f)$
and the corresponding pQCD values $a(Q^2;N_f)$, ${\rm rd}(Q^2) \equiv |\A_1(Q^2;N_f)/a(Q^2;N_f) - 1|$,
as a function of positive $Q^2$ and for various
$N_f$, is given in Fig.~\ref{reldiff}. 
These differences are extremely small, with the
exception of low $Q^2$: $0 < Q^2 < 1 \ {\rm GeV}^2$. 
When $N_f=4$, the difference $\A_1(Q^2;N_f)/a(Q^2;N_f) - 1$ changes
sign from negative to positive at increasing $Q^2$ around $Q^2 \approx 17 \ {\rm GeV}^2$; in the
case of $N_f=5$ this occurs around $Q^2 \approx 6 \ {\rm GeV}^2$. In the case of $N_f=3$ we have
$\A_1(Q^2;3)/a(Q^2;3) - 1 < 0$ for all positive $Q^2$.
These differences ${\rm rd}(Q^2)$ get smaller when $N_f$ increases. 
Therefore, the model 2$\delta$anQCD for $N_f \geq 4$
can be used in practical calculations of the underlying couplings
$a(Q^2)$ [$\approx \A_1(Q^2)$] and 
${\widetilde a}_{\nu}(Q^2)$ [$\approx \tA_{\nu}(Q^2)$]. We note that for any
real $\nu \geq 0$ we have
\be
\tA_{\nu}(Q^2;N_f) - {\widetilde a}_{\nu}(Q^2;N_f) \sim 
\left( \frac{\Lambda^2_{N_f}}{Q^2} \right)^5 \ ,
\label{diff2}
\ee
which is a consequence of Eq.~(\ref{ITEPc}). Namely, for integer $\nu=2,3,\ldots$ 
this can be obtained by applying $K_{\nu} (Q^2 d/d Q^2)^{\nu-1}$ to both sides of 
Eq.~(\ref{ITEPc}), where $K_{\nu} = (-1)^{\nu-1}/[\beta_0^{\nu-1} (\nu-1)!]$,
cf.~Eq.~(\ref{tAn}).\footnote{
We note that in such a case the derivative $(Q^2 d/d Q^2)^{\nu-1}$ applied to
$(\Lambda^2/Q^2)^5$ gives $(-5)^{\nu-1} (\Lambda^2/Q^2)^5$.}
And for $\nu$ noninteger Eq.~(\ref{diff2})
follows by analytic continuation of the integer case to $\nu$. We stress that
the exact calculation of the pQCD quantities ${\widetilde a}_{\nu}(Q^2;N_f)$
for noninteger $\nu$ is quite complicated, due to the Landau singularities
of the original pQCD coupling\footnote{
The coupling ${\widetilde a}_{\nu+1}(Q^2)$ for integer $\nu=n$ is a simple
$n$'th logarithmic derivative of $a(Q^2)$,
 ${\widetilde a}_{n+1}(Q^2) \equiv 
[(-1)^n/(\beta_0^n n!)] (\partial/\partial \ln Q^2)^n a(Q^2)$
[cf.~Eq.~(\ref{tAn})]. For noninteger $\nu$, ${\widetilde a}_{\nu+1}(Q^2)$
could be obtained by a dispersion integral similar to Eq.~(\ref{tAnu1}),
by including integration over the Landau cuts and poles ($\sigma < 0$).
This integration may be complicated, especially if an additional 
isolated Landau pole is involved as is the case of the coupling 
(\ref{aptexact}) with $c_2<0$ used here.}
$a(Q^2; N_f)$. Therefore, in the evaluations
of the series of the type
\bes
\label{ptmpt}
\bea
{\cal D}(Q^2) &=& a(Q^2)^{\nu_0} + \sum_{m=1}^{\infty} d_m a(Q^2)^{\nu_0 + m}
\label{pt}
\\
 &=& {\widetilde a}_{\nu_0}(Q^2) + \sum_{m=1}^{\infty} {\widetilde d}_m 
{\widetilde a}_{\nu_0 + m}(Q^2) 
\label{mpt}
\eea
\ees
with $\nu$ noninteger, the (truncated) expansion in the generalized logarithmic
derivatives (\ref{mpt}) can be evaluated in practice by applying the
model 2$\delta$anQCD (at a given $N_f$), 
as explained in Eqs.~(\ref{Dpt})-(\ref{Dan2}).
The (truncated) series in powers
(\ref{pt}) is, certainly, much easier to evaluate technically than the
(truncated) series (\ref{mpt}); nonetheless, the latter series may behave
in some cases better than the former, and then 2$\delta$anQCD can be
called upon, with the replacements: 
${\widetilde a}_{\nu_0+m}(Q^2;N_f) \mapsto \tA^{(2 \delta)}_{\nu_0+m}(Q^2;N_f)$ and
$a(Q^2;N_f)^{\nu_0+m} \mapsto \A^{(2 \delta)}_{\nu_0+m}(Q^2;N_f)$.
If the quantity ${\cal D}(Q^2)$ has low $Q^2$ corresponding to
$N_f=3$, the evaluation of the (truncated) series (\ref{mpt})
with the model 2$\delta$anQCD [${\widetilde a}_{\nu_0+m}(Q^2;3) \mapsto \tA^{(2 \delta)}_{\nu_0+m}(Q^2;3)$]  is then the natural and the 
preferred way of evaluation, because the (truncated) series (\ref{ptmpt})
in pQCD are usually numerically badly affected by the vicinity of Landau singularities
at such low $|Q^2| < (2 \m_c)^2$.

\subsection{Massive Perturbation Theory (MPT)}
 \label{sec:MPT}

In order to obtain a holomporphic coupling finite in the infrared regime,
the author of Ref.~\cite{MPT} proposed a simple change in the momentum
\be
\A_1^{\rm (MPT)}(Q^2; N_f) = a(Q^2+m_{\rm gl}^2; N_f) \ .
\label{MPT1}
\ee
The mass scale $m_{\rm gl} \approx 0.5-1$ GeV  is in this ansatz 
a constant and is associated  with an effective (dynamical) gluon mass
which reflects the infrared dynamics of QCD. The same kind of
replacement had been suggested, at one- and two-loop level, 
in Refs.~\cite{Simonov,BadKuz} as a result of
the use of nonperturbative QCD background. 
It was used in Refs.~\cite{BKS,KKSh} in analyses of
structure functions (with $m_{\rm gl} \approx 0.8$ GeV). The relation (\ref{MPT1}),
i.e., the replacement $Q^2 \mapsto Q^2+m_{\rm gl}^2$, can be kept
even at higher-loop levels, as suggested by the multiplicative renormalizability 
\cite{Luna} (and $m_{\rm gl}^2$ can be expected in general to run with $Q^2$). 
Such behavior is suggested also by Gribov-Zwanziger approach \cite{Gribov}, by
analyses of Dyson-Schwinger equations in QCD \cite{DSE1,DSE2}
and by other functional methods \cite{STQ,FRG}.

The coupling (\ref{MPT1}) is analytic, 
because $m^2_{\rm gl} > \Lambda^2_{\rm Lan.}$,
where $(-q^2 \equiv) Q^2=-\Lambda^2_{\rm Lan.}$ 
is the branching point of the Landau 
singularity cut of the corresponding pQCD coupling $a(Q^2)$. 
Therefore, $\A_1^{\rm (MPT)}(Q^2)$ can be written in the form (\ref{Adisp})
of dispersion integral, typical in any anQCD.
At large $|Q^2|$ the coupling $\A_1^{\rm (MPT)}(Q^2)$ tends to the
pQCD coupling $a(Q^2)$, the difference being
\be
\A_1^{\rm (MPT)}(Q^2; N_f) - a(Q^2; N_f)
\sim \frac{m_{\rm gl}^2}{Q^2 \ln^2(Q^2/{\overline \Lambda}^2)} \ .
\label{difmrho}
\ee
It is important to stress that, as  
$\A_1^{\rm (MPT)}(Q^2; N_f)$ is a nonperturbative
holomorphic coupling, the evaluation of the (truncated) perturbation power
series ${\cal D}^{[N]}(Q^2)$  of the spacelike scale- and scheme-invariant physical
quantities, Eq.~(\ref{Dpt}), should not be performed by replacing 
$a(\mu^2)^{\nu} \mapsto \A_1^{\rm (MPT)}(\mu^2)^{\nu}$, but by the replacement
which is obligatory in any anQCD
\be
 a(\mu^2)^{\nu} \mapsto \A_{\nu}(\mu^2) \ ,
\label{replMPT}
\ee
cf.~Eq.~(\ref{Dan}). The nonpower quantities 
$\A_{\nu}(\mu^2) = \A_{\nu}^{\rm (MPT)}(\mu^2)$ 
are constructed
via Eqs.~(\ref{AnutAnu}) and (\ref{tAnu2}), and in the integrands
of Eqs.~(\ref{tAnu2}) we use for $\A_1$ the expression (\ref{MPT1}).
This use of nonpower
expressions, based on the (generalized) logarithmic derivatives
$\tA_{\nu^{'}}(\mu^2)$ presented by Eq.~(\ref{tAnu1}) or Eq.~(\ref{tAnu2}), 
has been emphasized in Refs.~\cite{CV1,CV2,MPT}
for the case of integer $\nu$, extended to the case
of general (noninteger) $\nu$'s in Refs.~\cite{GCAK}, and
applied in various contexts in Refs.~\cite{Cvetic}.

Since for each given $N_f$ we have a specific underlying pQCD running
coupling $a(Q^2;N_f)$ in Eq.~(\ref{MPT1}), 
we have then the corresponding ${\rm MPT}_{N_f}$ model.
In general, $m_{\rm gl}$ may depend on $N_f$, as does the 
scale ${\overline {\Lambda}}_{N_f}$.

The generalized logarithmic derivatives $\tA_{\nu}$ are evaluated by
Eq.~(\ref{tAnu2}) for $0 \leq \nu < 5$, i.e., with $\nu=n+1+\delta$ where
$n+1=0,1,2,3,4$ and $0 \leq \delta < 1$.
We have $N$-loop ${\rm MPT}_{N_f}$ ($N=1,2,3,4$).
We call the model 1-loop ${\rm MPT}_{N_f}$ when
$a(Q^2;N_f)$ is 1-loop Eq.~(\ref{1LpQCD}) and in the construction
of $\A_{\nu}$ in Eq.~(\ref{AnutAnutr}) the right-hand side
has only one term: $\A_{\nu} = \tA_{\nu}$. We call the model
2-loop ${\rm MPT}_{N_f}$ when $a(Q^2;N_f)$ is 2-loop Eq.~(\ref{a2lLambert})
and in the construction of $\A_{\nu}$ in Eq.~(\ref{AnutAnutr}) the
right-hand side has two terms: $\A_{\nu} = \tA_{\nu} + {\tk}_1(\nu) \tA_{\nu+1}$
(except when $4 \leq \nu < 5$, in which case we take $\A_{\nu} = \tA_{\nu}$).
The model is called 3-loop ${\rm MPT}_{N_f}$ when $a(Q^2;N_f)$ is 
given by Eq.~(\ref{aptexact}) with $c_2={\overline c}_2(N_f)$ $\MSbar$ value
and in Eq.~(\ref{AnutAnutr}) the right-hand side has three terms:
$\tA_{\nu}= \tA_{\nu} + {\tk}_1(\nu) \tA_{\nu+1} + {\tk}_2(\nu) \tA_{\nu+2}$
(only two terms when $3 \leq \nu < 4$; only one term when $4 \leq \nu < 5$).
The model is called 4-loop ${\rm MPT}_{N_f}$ when $a(Q^2;N_f)$ is 
given by the expansion (\ref{NLpQCDapp}) with 
$c_2={\overline c}_2(N_f)$  and $c_3={\overline c}_3(N_f)$ 
(and $c_j=0$ for $j \geq 4$; ${\cal N}=8$ is used) and in 
Eq.~(\ref{AnutAnutr}) the right-hand side has in general four terms:
$\tA_{\nu}= \tA_{\nu} + \sum_{m=1}^3 {\tk}_m(\nu) \tA_{\nu+m}$
(only three terms when $2 \leq \nu < 3$; etc.).

If we take specific (input)
values of the dynamical masses $m_{\rm gl}(N_f)$ (for $N_f=3,4,5,6$),
and a specific value of  ${\overline {\Lambda}}_{3}$, the values
of other scales ${\overline {\Lambda}}_{N_f}$ (for $N_f=4,5,6$) can
be obtained by applying the quark threshold relations (\ref{msb})
written within MPT model
\begin{eqnarray}
\label{msbMPT}
\A_1^\prime &=&\A_1- \A_2 \frac{\ell_h}{6}
+ \A_3 \left(\frac{\ell_h^2}{36}-\frac{19}{24}\ell_h+ {\widetilde c}_2\right)
+ \A_4 \left[-\frac{\ell_h^3}{216}
\right.\nonumber\\
&-&\left.\vphantom{\frac{\ell_h^3}{216}}
\frac{131}{576}\ell_h^2+\frac{\ell_h}{1728} \left( -6793+281 (N_f-1) \right)
+ {\widetilde c}_3\right],
\end{eqnarray}
where $\A_1^\prime \equiv \A_1^{\rm (MPT)}(\mu^2_{N_f};N_f-1)$ and
$\A_n \equiv \A_n^{\rm (MPT)}(\mu^2_{N_f};N_f)$.

\subsection{Examples of various couplings as a function of positive $Q^2$}
 \label{sec:examples}

\begin{figure}[htb] 
\begin{minipage}[b]{.49\linewidth}
\centering\includegraphics[width=80mm]{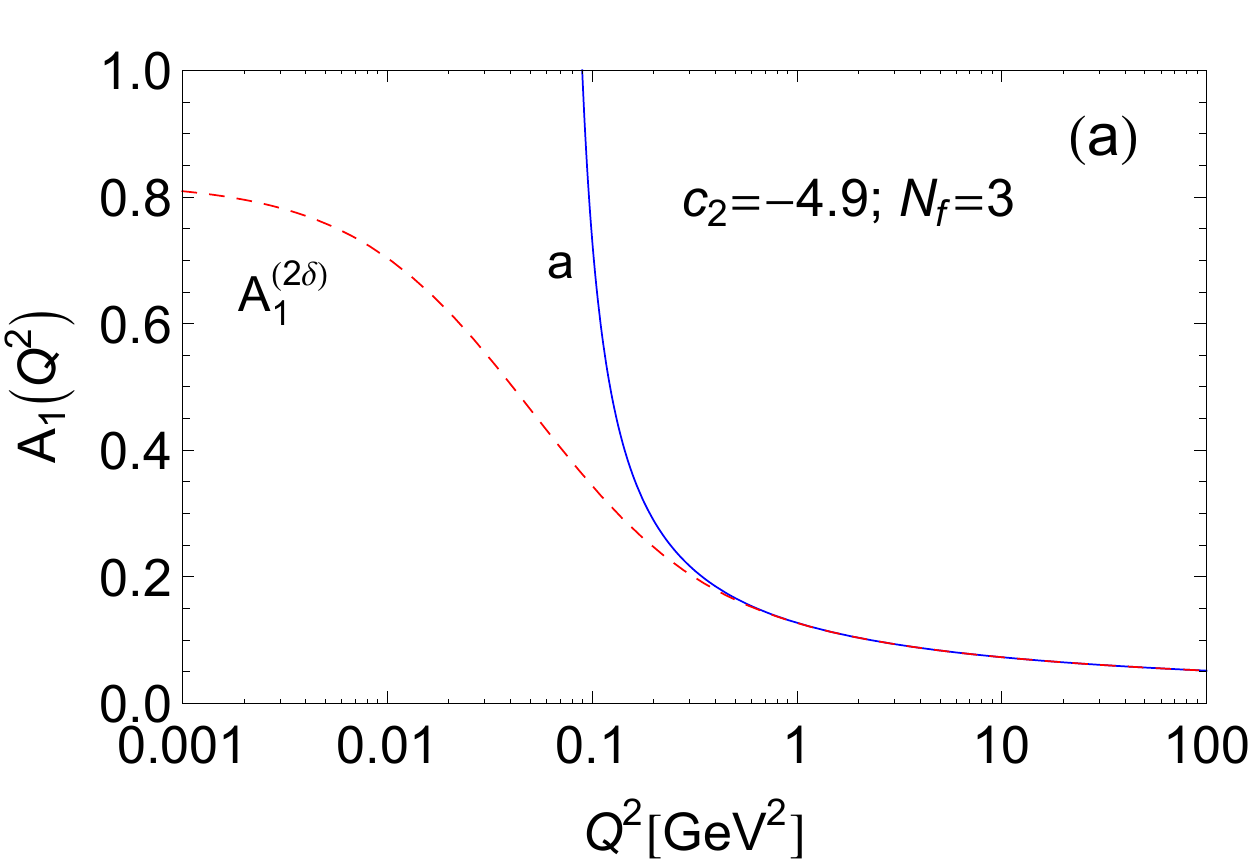}
\end{minipage}
\begin{minipage}[b]{.49\linewidth}
\centering\includegraphics[width=80mm]{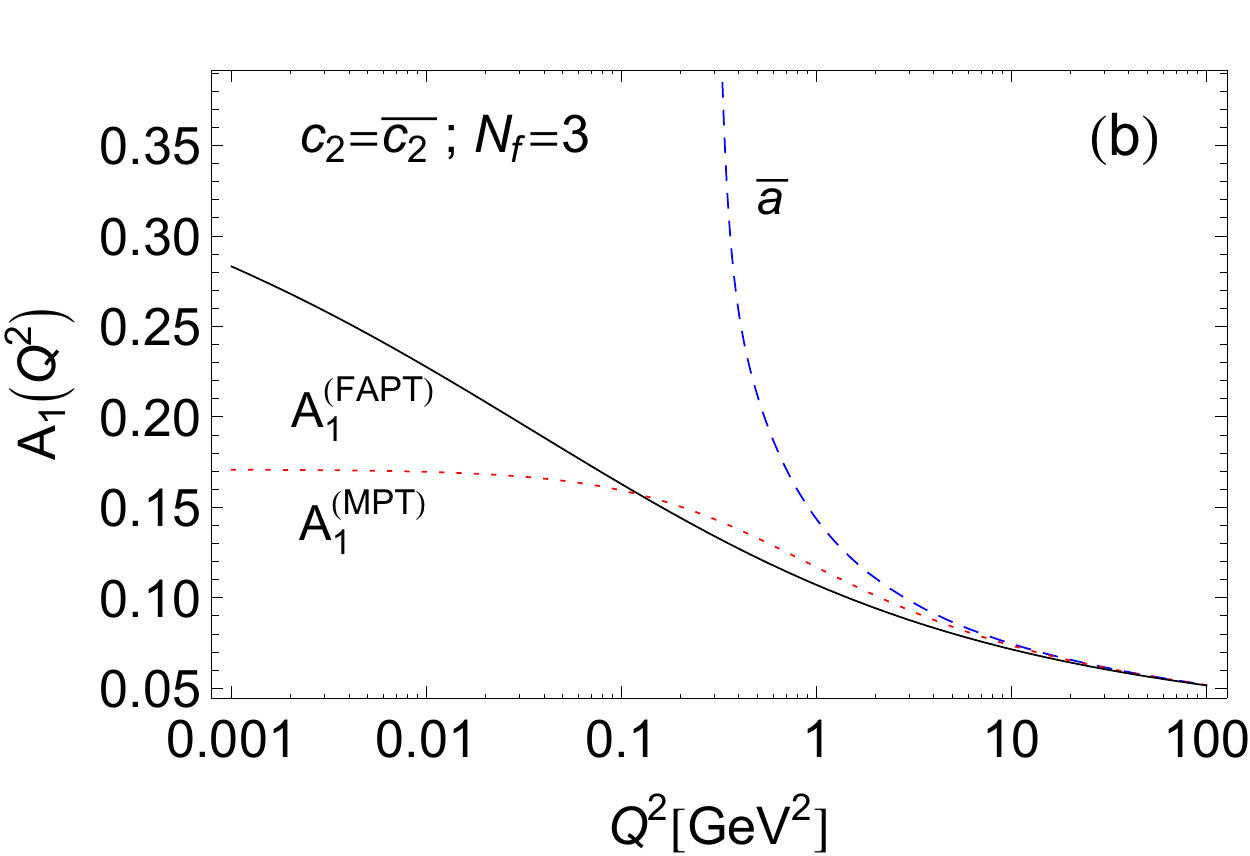}
\end{minipage}
\vspace{-0.4cm}
 \caption{\footnotesize  The couplings $\A_1 \equiv \A$ in 
three anQCD models with $\nu=1$ 
 and $N_f=3$ as a function of $Q^2$ (for $Q^2 >0$):
 (a) 2$\delta$anQCD coupling and pQCD coupling, in the renormalization
scheme with $c_2=-4.9$ (and $c_j=c_2^{j-1}/c_1^{j-2}$ for $j \geq 3$);
the underlying pQCD coupling $a$ is included for comparison;
 (b) FAPT and MPT in 4-loop $\MSbar$ scheme and with 
$\bL^2_3=0.1 \ {\rm GeV}^2$; MPT with $m^2_{\rm gl}=0.7 \ {\rm GeV}^2$;
${\overline a}$ is $a$ in $\MSbar$.}
\label{figA1L}
 \end{figure}

In Figs.~\ref{figA1L} we show the running of $\A_1(Q^2)$ for $Q^2>0$ and $N_f=3$
for three analytic models: FAPT, 2$\delta$anQCD, and MPT 
(with the choice $m^2_{\rm gl}=0.7 \ {\rm GeV}^2$). 
For comparison, we show also the underlying pQCD coupling $a(Q^2)$, i.e.,
$a(Q^2)$ in the same renormalization scheme and
with the same Lambert scale $\Lambda$. At low $Q^2$, the divergent behavior
of $a(Q^2)$ is evident, due to the Landau singularities.
We observe that at $Q^2 \gtrsim 1{\rm GeV}^2$ 2$\delta$anQCD coupling
is indistinguible from the underlying pQCD coupling, 
cf.~also Eq.~(\ref{ITEPc}).
FAPT and MPT anQCD couplings (presented here in 4-loop $\MSbar$ scheme)
are more suppressed in the infrared than 2$\delta$anQCD.
\begin{figure}[htb] 
\begin{minipage}[b]{.49\linewidth}
\centering\includegraphics[width=80mm]{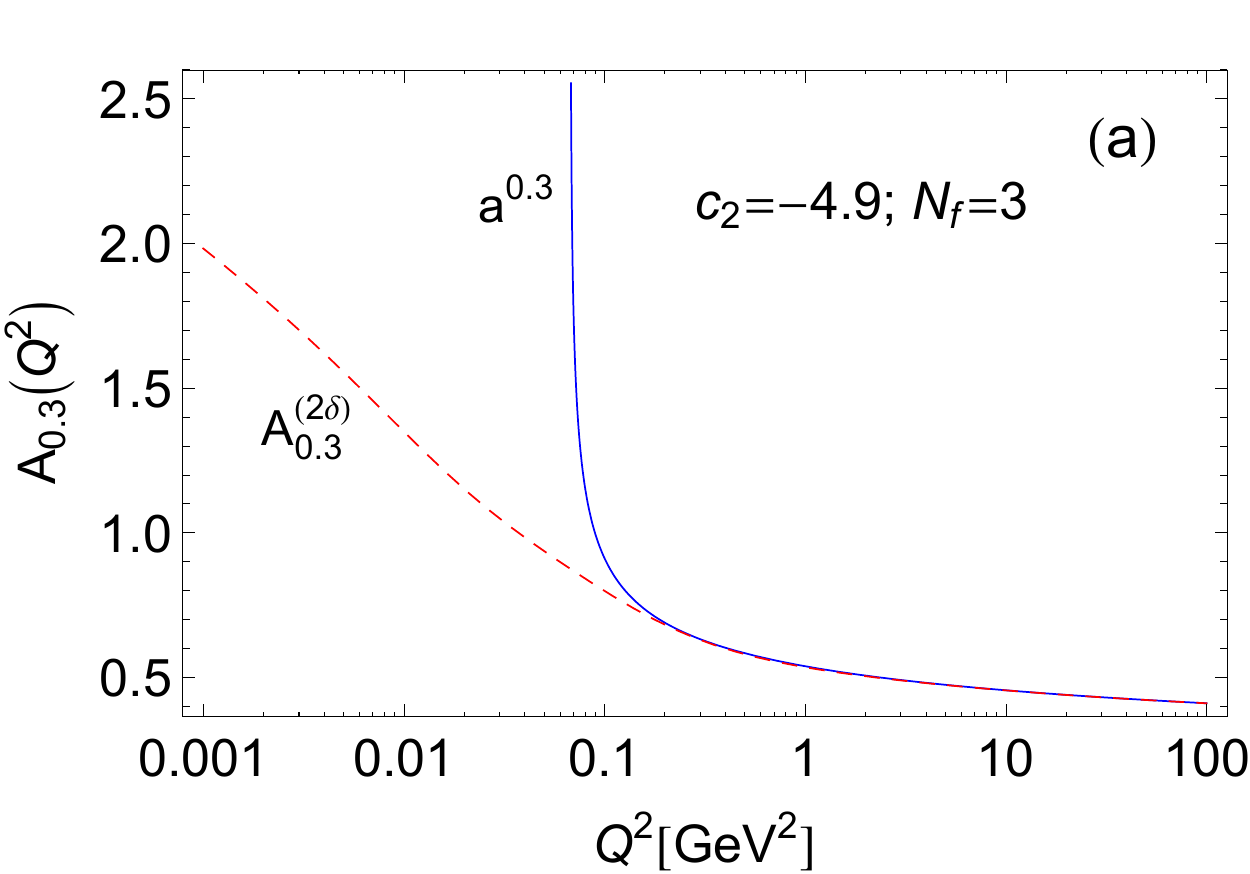}
\end{minipage}
\begin{minipage}[b]{.49\linewidth}
\centering\includegraphics[width=80mm]{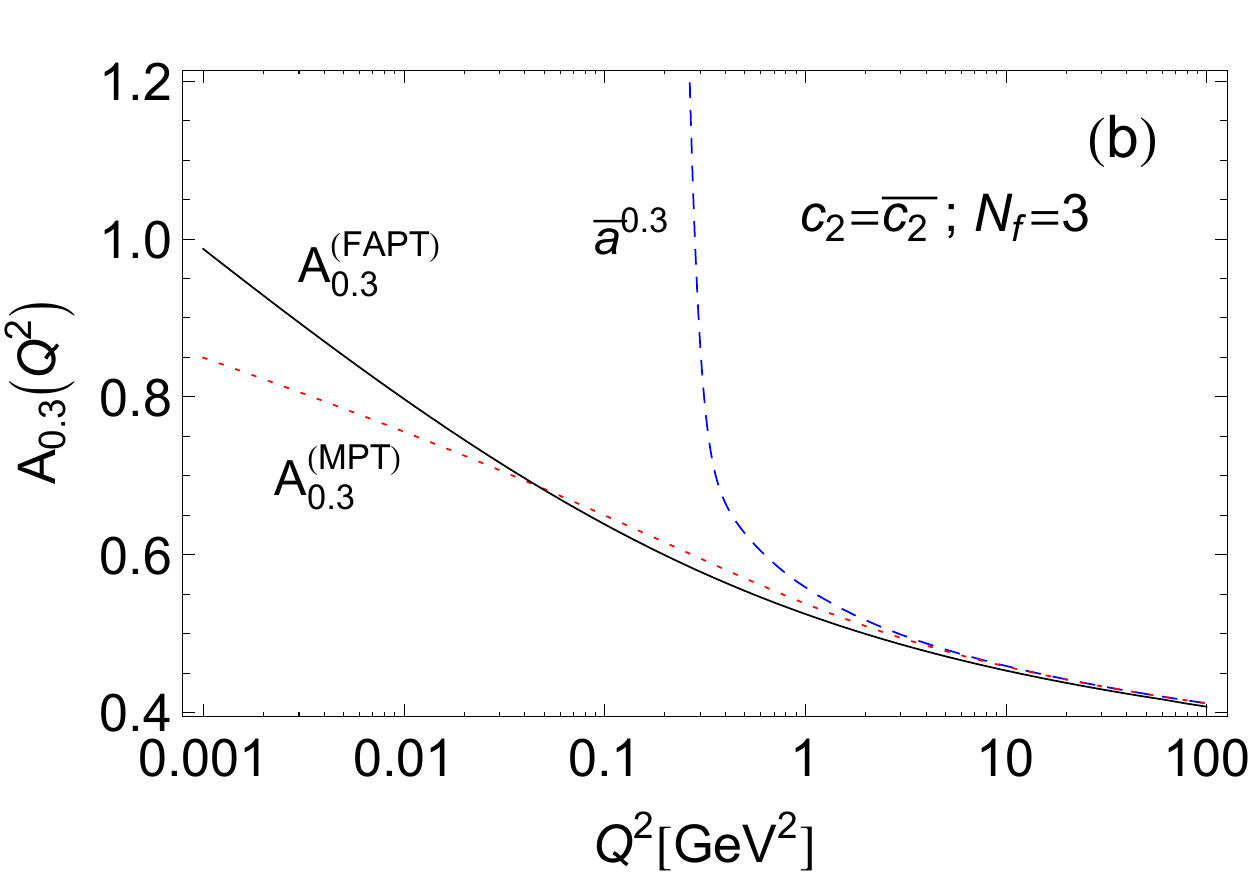}
\end{minipage}
\vspace{-0.4cm}
 \caption{\footnotesize The same as in Figs.~\ref{figA1L}, but 
 now with $\nu=0.3$ ($\A_{\nu=0.3}$). The coupling $\A_{0.3}$ is calculated
from the couplings $\tA_{0.3+m}$ using the relation (\ref{AnutAnutr}) 
(with $\nu_0=0.3$ and $n=0$) with the truncation index $N=5$ for 2$\delta$anQCD
and $N=4$ for MPT; and for FAPT using Eq.~(\ref{AnuFAPT}).}
\label{figA03L}
 \end{figure}
Figs.~\ref{figA03L} represent the couplings at $\nu=0.3$ (and $N_f=3$),
i.e., $\A_{\nu=0.3}(Q^2)$. We note the same behavior 
as in Figs.~\ref{figA1L}, but now MPT coupling increases more quickly 
when $Q^2$ decreases than in the $\nu=1$ case.

\section{Practical aspects of the program}
\label{sec:pract}

\subsection{Lambda scales and the treatment of quark thresholds}
\label{sec:Lthr}

We mention some practical aspects of the program, concerning the 
${\overline \Lambda}_{N_f}$ scales and the treatment of quark thresholds. 
The input parameter in the program is ${\overline \Lambda}_{N_f}^2$ 
(in ${\rm GeV}^2$) for fixed-$N_f$ FAPT and MPT 
models and  ${\overline \Lambda}_{3}^2$ for global FAPT.\footnote{
In global FAPT, the other ${\overline \Lambda}_{N_f}$ ($N_f>3$) are fixed
from   ${\overline \Lambda}_{3}$ by using for $a(Q^2)$ only the
expansion Eq.~(\ref{NLpQCDapp}) with ${\cal N}=8$ (and not the RGE-numerically
obtained ``exact'' values). But the effect of this additional approximation
in comparison to Table \ref{bLtab} in Sec.~\ref{sec:thr} is small. 
For example, for ${\overline \Lambda}_{3}=341.8$ MeV case
with $4/3$-loop approach and $\kappa=2$ (the first line in Table
\ref{bLtab}), the resulting  ${\overline \Lambda}_{N_f}$ becomes
$296.5$ MeV, $212.8$ MeV, $90.3$ MeV for $N_f=4,5,6$, respectively,
i.e., by about $0.5$ MeV lower than in Table \ref{bLtab}.
In $2/1$-loop approach with $\kappa=2$, 
for ${\overline \Lambda}_{3}=375.3$ MeV value (i.e., the second line of 
Table \ref{bLtab}), the values of ${\overline \Lambda}_{N_f}$ in
this approach are $311.9$ MeV, $215.8$ MeV and $89.4$ MeV
for  $N_f=4,5,6$, respectively, i.e., lower than in Table \ref{bLtab}
by less than $1$ MeV.}
In (fixed-$N_f$) 2$\delta$anQCD models, the scales
${\overline \Lambda}_{N_f}$ ($\Leftrightarrow \Lambda_{N_f}$ Lambert scales)
are fixed by the world average value $a(M_Z^2; \MSbar; N_f=5) = 0.1184/\pi$
\cite{PDG2012}. In addition, the scheme parameter 
$c_2$ ($\equiv \beta_2/\beta_0)$) in ${\rm 2}\delta{\rm anQCD}_{N_f=3}$ 
can be adjusted by hand and can vary in the interval 
$-5.6 < c_2 < -2$ (see later).
The quark threshold parameter is fixed to $\kappa=2$ in the 
program for 2$\delta$anQCD (kappa2d=2), and also in FAPT (kappa=2).
On the other hand, in MPT, at a given $N_f$, there is no $\kappa$
appearing, the scale ${\overline {\Lambda}}_{N_f}$ is an input parameter.
However, the value of $\kappa$ in global FAPT
can be adjusted by hand in the program,\footnote{Physically, $1 \leq \kappa \leq 3$ 
appears to be a reasonable interval of possible values. The values of various 
${\overline \Lambda}_{N_f}$ change very little when $\kappa$ is varied.}
while in 2$\delta$anQCD it should remain unchanged by construction (kappa2d=2).
If $N$ is the number of loops in the RGE running ($N=1,2,3$ or $4$), the input 
will be  ${\overline \Lambda}_{3}^2=$\texttt{L2}$N$\texttt{lnf3}
in global FAPT, and other 
scales (for other $N_f \equiv Nf$) are then given by the following functions: 
${\overline \Lambda}_{N_f}^{2}=
$\texttt{L2}$N$\texttt{l[}\textit{Nf,}\texttt{L2}N\texttt{lnf3]} 
with $Nf=4,5,6$ which is obtained via the $(N-1)$-loop matching condition, 
i.e., the relation (\ref{msb}) where, on the right-hand side, the last 
included term is $\sim a^N$.

Now, we consider an example of our pQCD running coupling and their value of Lambda QCD parameter, where the perturbative $N$-loop running coupling for $N_f$ is given by functions $(a1{\rm l}, a2{\rm l}, a3{\rm l}, a4{\rm l})$, where
\bea
 \label{aBar}
\text{a}N\text{l}[Nf,Q2,L2,\phi] & \equiv & 
a(Q^2=Q2 \times e^{i \phi}; N_f=Nf; L2={\overline \Lambda}_{Nf}^2; N{\rm -loop}; \MSbar) \ ,
\eea
where $Q2 = |Q^2|$, and $- \pi < \phi < \pi$.
The global running perturbative QCD coupling is
\bea
\text{a}N\text{lglob}[Nf,Q2,L23,\phi] & \equiv & 
a^{({\rm glob.})}(Q^2= Q2 \times e^{i \phi}; L23={\overline \Lambda}_{3 }^2; N{\rm -loop}; \MSbar) .
\label{aBarglob}
\eea
Our \texttt{Mathematica} package is called by the command

\begin{verbatim}
In[1]  :=  <<anQCD.m
\end{verbatim}

\noindent
Comment: We defined the physical parameters (mc$=\m_c$, etc.)
inside of the \texttt{NumDefanQCD} function:

\begin{verbatim}
In[2] := {mc/.NumDefanQCD, mb/.NumDefanQCD, mt/.NumDefanQCD, MZ/.NumDefanQCD}
Out[2] := {1.27, 4.2, 163., 91.1876}
\end{verbatim}
Comment: Lambda squared QCD parameter ${\overline {\Lambda}}_5^2$ can be fixed 
by the value $a(M_Z^2;\MSbar)=0.1184/\pi$
\begin{verbatim}
In[3] := L2nf5=L25/.FindRoot[a4l[5,91.1876^2/L25,0] == 0.1184/Pi,{L25,0.1}]
Out[3] := 0.0455164
\end{verbatim}

\subsection{Main procedures in analytic QCD models}
 \label{sec:proc}

We present here general rules on how to use the  \texttt{anQCD.m} package. 
For more detailed description we refer to \ref{sec:def}.
We present the main functions that we provide to the community: 
\begin{itemize}
\item $\text{tr}N\text{l}[N_f,\nu,k,\sigma,\bL^2_{N_f}]$ 
returns the $N$-loop perturbative spectral density 
$\rho_{\nu,k}^{(N)}(\sigma; N_f) = 
{\rm Im} \ [ a^{\nu} \ln^k a ]_{Q^2=-\sigma - i \epsilon}$ $(N=1, 2, 3, 4)$ of
real power $\nu$ and logarithmic power $k$ at $\sigma$
and at fixed number of active quark flavors $N_f$:
\bea
  \verb'tr'{N}\verb'l'[Nf,\nu,k,\sigma,L2]
 &=& \rho_{\nu,k}^{(N)} [ \sigma; Nf=N_f; L2=\bL^2_{N_f} ] 
\\
  && (\nu \in{\mathcal R}\,;\ k=0,1,\ldots\,;\ N = 1,2,3,4\,;\ Nf = 3,4,5,6).
\nonumber
\eea

\item  $\text{tr}N\text{lglob}[\nu,k,\sigma,\bL_3^2]$ returns the $N$-loop 
global perturbative spectral density 
$\rho_{\nu,k}^{(N){\rm glob.}}(\sigma; N_f)$ $(N=1, 2, 3, 4)$ of
real power $\nu$ and logarithmic power $k$ at $\sigma$, 
and with $\bL_3$ being the QCD $N_f=3$ scale:

\be
  \verb'tr'{N}\verb'lglob'[\nu,k,\sigma,L23]
   =\rho_{\nu,k}^{(N){\rm glob.}}[\sigma; L23=\bL_3^2]\,,\quad
  (N = 1,2,3,4).
\ee

\item $\texttt{AFAPT}N\texttt{l}[N_f,\nu,k,|Q^2|,\Lambda^2,\phi]$
returns the $N$-loop $(N=1, 2, 3, 4)$
analytic FAPT coupling  $\A_{\nu,k}^{({\rm FAPT},N)}(Q^2, N_f) =
( a^{\nu}(Q^2) \ln^k a(Q^2) )_{\rm an.FAPT}$,
of real power $\nu$
and logarithmic power $k$ at fixed number of active quark 
flavors $N_f$, 
in the Euclidean domain [$Q^2 = |Q^2| \exp(i \phi) 
\in {\mathcal C}$ and $Q^2 \not< 0$], with $Q^2$ in units of ${\rm GeV}^2$
and $\phi$ in radians
\bea
  \verb'AFAPT'{N}\verb'l'[Nf,\nu,k,Q2,L2,\phi] &= & 
\nonumber\\
= {\A}_{\nu,k}^{({\rm FAPT},N)}[Q2=|Q^2|, \phi={\rm arg}(Q^2); 
Nf=N_f; L2=\bL^2_{N_f}] &&
  \nonumber\\
  \quad (N = 1,2,3,4\,;\ Nf = 3,4,5,6). &&
\eea

\item In the global FAPT case
$\texttt{AFAPT}N\texttt{lglob}[\nu,k,|Q^2|,\Lambda_3^2,\phi]$
returns the $N$-loop analytic FAPT coupling 
${\A}_{\nu,k}^{({\rm FAPT},N){\rm glob.}}(Q^2)$.
of real power $\nu$ 
and logarithmic power $k$, in the Euclidean domain,
\bea
  \verb'AFAPT'{N}\verb'lglob'[\nu,k,Q2,L23,\phi]&=&
  {\A}_{\nu,k}^{({\rm FAPT},N){\rm glob.}}[Q2\!=|Q^2|, \phi\!={\rm arg}(Q^2); 
L23\!=\bL_3^2]
\nonumber\\
  && (N = 1,2,3,4).
\eea

\item $ \texttt{tA2d}[N_f,\nu,|Q^2|,\phi]$
returns the analytic 2$\delta$anQCD coupling 
${\widetilde {\A}}_{\nu}^{(2\delta)}(Q^2, N_f)$, 
the generalized logarithmic derivative
with index $\nu$ ($\nu > -1$ and real, in general noninteger),
at fixed number of active quark flavors $N_f$,
in the Euclidean domain $Q^2 = |Q^2| \exp(i \phi) 
\in {\mathcal C} \backslash [-M^2_{\rm thr.},-\infty)$ where
$M_{\rm thr.}^2 = M_2^2$ ($= s2s0[N_f] LL2[N_f]$)
\bea
  \verb'tA2d'[Nf,\nu,Q2,\phi]
  &=& {\widetilde {\A}}_{\nu}^{(2\delta)}[Q2=|Q^2|, \phi={\rm arg}(Q^2); Nf= N_f]\,,
  \nonumber\\
  &&  \!\!\!\!\!\!\!\! \!\!\
(Nf= 3,4,5,6;\ \nu>-1).
\eea

\item $ \texttt{A2d}N\texttt{l}[N_f,n,\nu,|Q^2|,\phi]$
returns the $N$-loop analytic 2$\delta$anQCD coupling 
${\A}_{n+\nu}^{(2\delta)}(Q^2, N_f)$, 
of fractional power $n+\nu$ ($\nu > -1$ and real; $n=0,1,\ldots,N-1$)
at fixed number of active quark flavors $N_f$,
in the Euclidean domain $Q^2 = |Q^2| \exp(i \phi) 
\in {\mathcal C} \backslash [-M^2_{\rm thr.},-\infty)$ where
$M_{\rm thr.}^2 = M_2^2$, 
used for the ${\rm N}^{N-1}{\rm LO}$
truncation approach [cf.~Eqs.~(\ref{Dpt})-(\ref{td}), in particular
Eq.~(\ref{AnutAnutr}) with $\nu \mapsto \nu_0$]
\bea
  \verb'A2d'{N}\verb'l'[Nf,n,\nu,Q2,\phi]
  &=& {\A}_{\nu+n}^{(2\delta)}[Q2=|Q^2|, \phi={\rm arg}(Q^2); Nf= N_f]\,,
  \nonumber\\
  &&  \!\!\!\!\!\!\!\! \!\!\!\!\!\!\!\!\!\!\!\!\!\!\!\! \!\!\!\!\!\!\!
(N= 1,2,3,4,5;\ Nf= 3,4,5,6;\ n=0,1,\ldots,N-1).
\eea

\item $\texttt{tAMPT}N\texttt{l}[N_f,n,\nu,Q^2,m_{\rm gl}^2,\bL^2_{N_f}]$
returns the $N$-loop $(N=1, 2, 3, 4)$
analytic MPT coupling 
${\widetilde {\A}}_{n+\nu}^{({\rm MPT},N)}(Q^2, m_{\rm gl}^2,N_f)$,
the generalized logarithmic derivative
with index $n+\nu$ ($n=0,1,2,3,4; 0 \leq \nu<1$),
at fixed number of active quark flavors $N_f$,
with $Q^2$ in the Euclidean domain ($Q^2 \in {\mathcal C}$ and $Q^2 \not< 0$)
\bea
  \verb'tAMPT'{N}\verb'l'[Nf,n,\nu,Q2,M2,L2] = &&
\nonumber\\
  = {\widetilde {\A}}_{n+\nu}^{(\rm{MPT},N)}[Q2=Q^2 \in {\mathcal C}; Nf=N_f; M2=m_{\rm gl}^2; L2=\bL^2_{N_f}] &&
  \nonumber\\
 \!\!\!\!\!\!\!\! \!\!\!\!\!\!\!\!\!\!\!\!\!\!\!\! \!\!\!\!\!\!\!\!\!\!\!\!
(N = 1,2,3,4\,;\ Nf = 3,4,5,6)\,;\ n=0,1,2,3,4;\ 0 \leq \nu < 1). &&
\eea

\item $\texttt{AMPT}N\texttt{l}[N_f,\nu,Q^2,m_{\rm gl}^2,\bL^2_{N_f}]$
returns the $N$-loop $(N=1, 2, 3, 4)$
analytic MPT coupling 
${\A}_{\nu}^{({\rm MPT},N)}(Q^2, m_{\rm gl}^2,N_f)$,
of fractional power $\nu$ ($0<\nu<5$)
and at fixed number of active quark flavors $N_f$,
with $Q^2$ in the Euclidean domain ($Q^2 \in {\mathcal C}$ and $Q^2 \not< 0$)
\bea
  \verb'AMPT'{N}\verb'l'[Nf,\nu,Q2,M2,L2] = &&
\nonumber\\
  = {\A}_{\nu}^{(\rm{MPT},N)}[Q2=Q^2 \in {\mathcal C}; Nf=N_f; M2=m_{\rm gl}^2; L2=\bL^2_{N_f}] &&
  \nonumber\\
(N = 1,2,3,4\,;\ Nf = 3,4,5,6)\,;\ 0 < \nu < 5). &&
\eea
\end{itemize}   

\subsection{Examples of the use}
 \label{sec:ex}

With the main procedures and definitions given above, 
we provide a few examples of the use of these quantities for 
\texttt{Mathematica}~9.0.1 and \texttt{Mathematica}~10.0.1. 
\begin{verbatim}
In[1]:= <<anQCD.m;
\end{verbatim}
We illustrate now how to obtain the values of the 
analytic couplings at what we call the three-loop level ($N=3$), i.e.,
the underlying pQCD coupling is given by 
Eq.~(\ref{aptexact}) with $c_2=c_2(N_f;\MSbar)$
in FAPT and MPT, and $c_2=-4.9$ in 2$\delta$anQCD.
Thus, we evaluate
${\A}_{\nu,0}^{({\rm FAPT},N)}(Q^2)$,
${\A}_{\nu,0}^{({\rm FAPT},N){\rm glob.}}(Q^2)$, ${\tA}_{\nu}^{(2\delta)}(Q^2)$, 
${\A}_{n+\nu}^{(2\delta)}(Q^2)$, ${\tA}_{n+\nu}^{(\rm{MPT},N)}(Q^2)$  and 
${\A}_{\nu}^{(\rm{MPT},N)}(Q^2)$, taking the 
parameters: $\bL_3^2=0.1 \ {\rm GeV}^2$ (in FAPT and MPT); 
$m_{\rm gl}^2 = 0.7 \ {\rm GeV}^2$ in MPT.
For the momentum scales we take $Q^2=10^{-3} \ {\rm GeV}^2$ (and $N_f=3$);
 $Q^2=10^{2} \ {\rm GeV}^2$ (and $N_f=5$);
$Q^2 =0.5 \times \exp(i 0.9) \ {\rm GeV}^2$ (and $N_f=3$).
We employ the indices $\nu=1$; $\nu=1.4$ ($n=1$ and $\nu=0.4$).
The calculated values of the couplings are given below 
(as the {\it second} entry), 
with the corresponding typical calculation time in seconds
(as the first entry, varies with various computers):\footnote{
The typical times are given when \texttt{Mathematica}~9.0.1 is used. When
\texttt{Mathematica}~10.0.1 is used, the times are in general longer by
about $20$-$50\%$.}
\begin{verbatim}
In[2]:= AFAPT3l[3, 1, 0, 10^-3, 0.1, 0] // Timing
Out[2]= {0.404938, 0.28312}
\end{verbatim}
\begin{verbatim}
In[3]:= AFAPT3lglob[1, 0, 10^-3, 0.1, 0] // Timing
Out[3]= {0.822874, 0.287775}
\end{verbatim}
\begin{verbatim}
In[4]:= A2d3l[3, 0, 1, 10^-3, 0] // Timing
Out[4]= {0.386942, 0.809041}
\end{verbatim}
\begin{verbatim}
In[5]:= AMPT3l[3, 1, 10^-3, 0.7, 0.1] // Timing
Out[5]= {0.150978, 0.171356}
\end{verbatim}
\begin{verbatim}
In[6]:= AFAPT3l[5, 1, 0, 10^2, 0.1, 0] // Timing
Out[6]= {0.410938, 0.0624843}
\end{verbatim}
\begin{verbatim}
In[7]:= AFAPT3lglob[1, 0, 10^2, 0.1, 0] // Timing
Out[7]= {0.809877, 0.0559854}
\end{verbatim}
\begin{verbatim}
In[8]:= A2d3l[5, 0, 1, 10^2, 0] // Timing
Out[8]= {0.510922, 0.0559197}
\end{verbatim}
\begin{verbatim}
In[9]:= AMPT3l[5, 1, 10^2, 0.7, 0.1] // Timing
Out[9]= {0.115982, 0.0627726}
\end{verbatim}
\begin{verbatim}
In[10]:= AFAPT3l[3, 1.4, 0, 0.5, 0.1, 0.9] // Timing
Out[10]= {0.400939, 0.0458667 - 0.00873018 I}
\end{verbatim}
\begin{verbatim}
In[11]:= AFAPT3lglob[1.4, 0, 0.5, 0.1, 0.9] // Timing
Out[11]= {0.861869, 0.0480877 - 0.00873811 I}
\end{verbatim}
\begin{verbatim}
In[12]:= tA2d[3, 1.4, 0.5, 0.9] // Timing 
Out[12]= {0.763884, 0.0694758 - 0.0380018 I} 
\end{verbatim}
\begin{verbatim}
In[13]:= A2d3l[3, 1, 0.4, 0.5, 0.9] // Timing
Out[13]= {1.543767, 0.062836 - 0.0325823 I}
\end{verbatim}
\begin{verbatim}
In[14]:= tAMPT3l[3, 1, 0.4, 0.5 Exp[I 0.9], 0.7, 0.1] // Timing 
Out[14]= {0.049993, 0.0555028 - 0.00617486 I} 
\end{verbatim}
\begin{verbatim}
In[15]:= AMPT3l[3, 1.4, 0.5 Exp[I 0.9], 0.7, 0.1] // Timing
Out[15]= {0.175973, 0.0537096 - 0.00719995 I}
\end{verbatim}

In order to make plots of the analytic running couplings as in 
Fig.~\ref{figA1L} and \ref{figA03L}, users could construct an 
interpolation in order to reduce the time of calculation.

\vspace*{+7mm}
\textbf{Acknowledgments}\vspace*{+1mm}
This work was supported by FONDECYT (Chile) Grant No. 1130599 and DGIP 
(UTFSM) internal project USM No. 11.13.12 (C.A and G.C). 

\vspace*{+7mm}

\begin{appendix}
\appendix

\vspace*{7mm}
\section{Description of the main procedures}
\label{sec:def}

The main functions found in our package are presented 
and described in the following.

\begin{flushleft}
\begin{itemize}
 \item \texttt{tr}$N$\texttt{l[Nf,Nu,k,sig,L2]}:
  \begin{itemize}
  \item[\textit{general:}] it computes the $N$-loop spectral density including
possibly powers of the logarithmic coupling,
   $\rho_{\nu,k}^{(N)}(\sigma, N_f)=
{\rm Im}[a(Q^2)^{\nu}{\rm ln}^k(a(Q^2))]_{Q^2=-\sigma-i \epsilon}$;
  \item[\textit{input:}] 
   the number of active flavors \texttt{Nf}=$N_f$;
   the power index \texttt{Nu}=$\nu$ and the logarithmic power index \texttt{k}=$k$;
the squared momentum argument \texttt{sig}=$\sigma$;
  the squared $\MSbar$ Lambda QCD parameter \texttt{L2}=$\overline{\Lambda}_{N_f}^2$ (all scales in ${\rm GeV}^2$);
  \item[\textit{output:}] $\rho^{(N)}_{\nu,k}$;
  \item[\textit{example:}] In order to compute the value of
    the three-loop spectral density, at $\sigma = 1.5 \ {\rm GeV}^2$ and $N_f=3$,
and with $\bL^2_{N_f}=0.1 \ {\rm GeV}^2$, i.e.,
  the quantity
    $\rho_{0.5,0}^{(3)}(1.5, 3)=0.104393$,
    one has to use the command
    \verb|tr3l[3,0.5,0,1.5,0.1]|.
  \end{itemize}

 \item \texttt{tr}$N$\texttt{lglob[Nu,k,sig,L2nf3]}:
  \begin{itemize}
  \item[\textit{general:}] it computes the $N$-loop global spectral density incorporating 
  the powers of the logarithmic coupling 
   $\rho_{\nu,k}^{(N){\rm glob.}}(\sigma, N_f)={\rm Im}[a^{\rm (glob.)}(Q^2)^{\nu}{\rm ln}^k(a^{\rm (glob.)}(Q^2))]_{Q^2=-\sigma - i \epsilon}$;
  \item[\textit{input:}] 
   the power index \texttt{Nu}=$\nu$ and the logarithmic power index \texttt{k}=$k$;
the squared momentum argument \texttt{sig}=$\sigma$;
  the squared $\MSbar$ Lambda QCD parameter at $N_f=3$ (at the corresponding $N$-loop) \texttt{L2nf3}=$\overline{\Lambda}_3^2$ (all scales are in ${\rm GeV}^2$);
  \item[\textit{output:}] $\rho^{(N){\rm glob.}}_{\nu,k}$;
  \item[\textit{example:}] In order to compute the value of
    the three-loop global spectral density at $\sigma = 1.5 \ {\rm GeV}^2$
and with $\bL^2_{3}=0.1 \ {\rm GeV}^2$, i.e.,
  the quantity $\rho_{0.5,0}^{(3){\rm glob.}}(1.5, 3)=0.104393$,
    one has to use the command
    \verb|tr3lglob[0.5, 0, 1.5, 0.1]|.
  \end{itemize}

 \item \texttt{AFAPT}$N$\texttt{l[Nf,Nu,k,Q2,L2,Fi]}:
  \begin{itemize}
  \item[\textit{general:}] it computes the $N$-loop coupling in ${\rm FAPT}_{N_f}$ 
  incorporating the analytization of powers of the logarithmic coupling
   $\A_{\nu,k}^{({\rm FAPT},N)}(Q^2, N_f) =
( a^{\nu}(Q^2) \ln^k a(Q^2) )_{\rm an.FAPT}$ in the Euclidean domain;
  \item[\textit{input:}] 
   the number of active flavors \texttt{Nf}=$N_f$;
   the power index \texttt{Nu}=$\nu$ and the logarithmic power index \texttt{k}=$k$;
the squared momentum argument \texttt{Q2}=$|Q^2|$;
  the squared $\MSbar$ Lambda QCD parameter \texttt{L2}=$\overline{\Lambda}_{N_f}^2$;
the phase of the complex $Q^2 = |Q^2| e^{i \phi}$, i.e., \texttt{Fi}=$\phi$ (in radians); all scales are in ${\rm GeV}^2$;
  \item[\textit{output:}] ${\A}_{\nu,k}^{({\rm FAPT},N)}$;
  \item[\textit{example:}] In order to compute the value of
    the three-loop FAPT coupling
 $\A_{\nu}$ at $Q^2=1.5 \ {\rm GeV}^2$, 
 with $\nu=0.5$, $N_f=3$ and $\bL^2_3 = 0.1 \ {\rm GeV}^2$, i.e., the quantity
    ${\A}_{0.5,0}^{({\rm FAPT},3)}(1.5,3)=0.324597$,
    one has to use the command
    \verb|AFAPT3l[3, 0.5, 0, 1.5, 0.1, 0]|.
  \end{itemize}

 \item \texttt{AFAPT}$N$\texttt{lglob[Nu,k,Q2,L2nf3,Fi]}:
  \begin{itemize}
  \item[\textit{general:}] it computes the $N$-loop global FAPT coupling
${\A}_{\nu,k}^{({\rm FAPT},N){\rm glob.}}(Q^2)$ in the Euclidean domain;
  \item[\textit{input:}] 
   the power index \texttt{Nu}=$\nu$ and the logarithmic power index \texttt{k}=$k$;
the squared momentum argument \texttt{Q2}=$|Q^2|$;
  the squared $\MSbar$ Lambda QCD parameter at $N_f=3$
  \texttt{L2nf3}=$\bL_3^2$;
 the phase of the complex $Q^2 = |Q^2| e^{i \phi}$, i.e., \texttt{Fi}=$\phi$ (in radians); all scales are in ${\rm GeV}^2$;
  \item[\textit{output:}] ${\A}_{\nu,k}^{({\rm FAPT},N){\rm glob.}}$;
  \item[\textit{example:}] In order to compute the value of
    the three-loop FAPT coupling $\A_{\nu}$ at $Q^2=1.5 \ {\rm GeV}^2$, 
 with $\nu=0.5$ and $\bL^2_3 = 0.1 \ {\rm GeV}^2$, i.e., the quantity
    ${\A}_{0.5,0}^{({\rm FAPT},3){\rm glob.}}(1.5)=0.333458$,
    one has to use the command
    \verb|AFAPT3lglob[0.5, 0, 1.5, 0.1, 0]|.
  \end{itemize}

 \item \texttt{tA2d}\texttt{[Nf,nu,Q2,Fi]}: 
  \begin{itemize}
  \item[\textit{general:}] it computes coupling
   ${\widetilde {\A}}_{\texttt{nu}}^{(2\delta)}(Q^2, N_f)$ in ${\rm 2}\delta{\rm anQCD}_{N_f}$, the generalized logarithmic derivative with index 
\texttt{nu}, in the Euclidean domain;
  \item[\textit{input:}] the number of active flavors \texttt{Nf}=$N_f$;
   the index $\nu=\texttt{nu}$ ($\texttt{nu} > -1$ and real); 
the squared momentum argument \texttt{Q2}=$|Q^2|$ (in ${\rm GeV}^2$);
 \texttt{Fi}=$\phi$ is the phase of the complex $Q^2 = |Q^2| e^{i \phi}$
(in radians); 
  \item[\textit{output:}] ${\widetilde {\A}}_{\nu}^{(2\delta,N)}$;
  \item[\textit{example:}] In order to compute the value of
    ${\widetilde {\A}}_{\nu}$ at $Q^2=0.5 \ {\rm GeV}^2$, 
 with $\texttt{nu}=1.4$, and $N_f=3$, i.e., the coupling
     ${\widetilde {\A}}_{1.4}^{(2\delta,3)}(0.5)=0.0827052$,
 one has to use the command
    \verb|tA2d[3, 1.4, 0.5, 0]|. 

  \end{itemize}
 
 \item \texttt{A2d}$N$\texttt{l[Nf,n,nu,Q2,Fi]}: 
  \begin{itemize}
  \item[\textit{general:}] it computes $N$-loop coupling
   ${\A}_{\texttt{nu}+\texttt{n}}^{(2\delta)}(Q^2, N_f)$ in ${\rm 2}\delta{\rm anQCD}_{N_f}$
in the Euclidean domain;
  \item[\textit{input:}] the number of active flavors \texttt{Nf}=$N_f$;
the indices $\texttt{n}$ ($\texttt{n}$ is nonnegative integer) and 
$\texttt{nu}$ ($\texttt{nu}> -1$ and real); 
the squared momentum argument \texttt{Q2}=$|Q^2|$ (in ${\rm GeV}^2$),
 \texttt{Fi}=$\phi$ is the phase of the complex $Q^2 = |Q^2| e^{i \phi}$
(in radians); 
see also Eq.~(\ref{AnutAnutr}), with
$\nu_0 \mapsto \texttt{nu}$ and $n \mapsto \texttt{n}$;
  \item[\textit{output:}] ${\A}_{\nu+n}^{(2\delta,N)}$;
  \item[\textit{example:}] In order to compute the value of
    the ``three-loop'' 2danQCD coupling $\A_{\nu}$ at $Q^2=0.5 \ {\rm GeV}^2$, 
 with $\texttt{nu}=0.4$, $\texttt{n}=1$ and $N_f=3$, i.e., the coupling
     ${\A}_{1.4}^{(2\delta,3)}(0.5)=0.0745576$,
 one has to use the command
    \verb|A2d3l[3, 1, 0.4, 0.5, 0]|. 

  \end{itemize}

 \item \texttt{tAMPT}$N$\texttt{l[Nf,n,nu,Q2,M2,L2MPT]}:
  \begin{itemize}
  \item[\textit{general:}] it computes the coupling
   ${\widetilde {\A}}_{\texttt{n}+\texttt{nu}}^{({\rm MPT},N)}(Q^2, m_{\rm gl}^2,N_f)$,
the generalized logarithmic derivative with index 
$\texttt{n}+\texttt{nu}$,
in ${\rm MPT}_{N_f}$ in the Euclidean domain;
  \item[\textit{input:}] 
   the number of active flavors $\texttt{Nf}=N_f$; 
   the integer index $\texttt{n}$ ($=0,1,2,3,4$) and the 
noninteger index \texttt{nu}=$\nu$ ($0 \leq \nu < 1$);
the squared momentum argument \texttt{Q2}=$Q^2$ (complex
in general);
   the effective mass parameter \texttt{M2}=$m_{\rm gl}^2$;
  the squared $\MSbar$ Lambda QCD parameter \texttt{L2MPT}=$\overline{\Lambda}_{N_f}^2$ (all scales in ${\rm GeV}^2$); all scales are in ${\rm GeV}^2$;
  \item[\textit{output:}] ${\widetilde {\A}}_{n+\nu}^{({\rm MPT},N)}$;
  \item[\textit{example:}] In order to compute the value of
    the three-loop MPT coupling ${\widetilde {\A}}_{n+\nu}$ with $N_f=3$,
 with $n=1$ and $\nu=0.4$, at $Q^2=0.5 \ {\rm GeV}^2$, 
 with $m_{\rm gl}^2=0.7 \ {\rm GeV}^2$, and $\bL_3^2=0.1 \ {\rm GeV}^2$, 
i.e., the quantity
    ${\widetilde {\A}}_{1.4}^{({\rm MPT},3)}(0.5,0.7,3)=0.0528178$,
    one has to use the command
    \verb|tAMPT3l[3, 1, 0.4, 0.5, 0.7, 0.1]|.
  \end{itemize}

 \item \texttt{AMPT}$N$\texttt{l[Nf,Nu,Q2,M2,L2MPT]}:
  \begin{itemize}
  \item[\textit{general:}] it computes the $N$-loop coupling
   ${\A}_{\nu}^{({\rm MPT},N)}(Q^2, m_{\rm gl}^2,N_f)$ 
in ${\rm MPT}_{N_f}$ in the Euclidean domain;
  \item[\textit{input:}] 
   the number of active flavors \texttt{Nf}=$N_f$;
   the index \texttt{Nu}=$\nu$ ($0 < \nu < 5$);
the squared momentum argument \texttt{Q2}=$Q^2$ (complex
in general);
  the squared $\MSbar$ Lambda QCD parameter \texttt{L2MPT}=$\overline{\Lambda}_{N_f}^2$;
   the effective mass parameter \texttt{M2}=$m_{\rm gl}^2$
(all scales in ${\rm GeV}^2$);
  \item[\textit{output:}] ${\A}_{\nu}^{({\rm MPT},N)}$;
  \item[\textit{example:}] In order to compute the value of
    the three-loop MPT coupling $\A_{\nu}$ with $N_f=3$,  $\nu=1.4$,
at $Q^2=0.5 \ {\rm GeV}^2$, with $m_{\rm gl}^2=0.7 \ {\rm GeV}^2$, 
and $\bL_3^2=0.1 \ {\rm GeV}^2$, 
i.e., the quantity
    ${\A}_{1.4}^{({\rm MPT},3)}(0.5,0.7,3)=0.0514469$,
    one has to use the command
    \verb|AMPT3l[3, 1.4, 0.5, 0.7, 0.1]|.
  \end{itemize}
\end{itemize}
\end{flushleft}

All scales ${\overline \Lambda}^2_{N_f}$, $Q^2$ (Euclidean),
and spectral-integration variables $\sigma$ are in GeV$^2$.
The number of loops $N$ is specified 
in the names of the procedures, except in 2$\delta$anQCD where
the underlying pQCD coupling is given by Eq.~(\ref{aptexact})
with $c_2=-4.9$ (this value can be changed by hand in the program
anQCD.m, by replacing ``c22din=-4.9;'' by another value, between -5.6 and -2.0).
\end{appendix}


\newcommand{\noopsort}[1]{} \newcommand{\printfirst}[2]{#1}
 \newcommand{\singleletter}[1]{#1} \newcommand{\switchargs}[2]{#2#1}

\end{document}